\documentclass[floatfix,prd,twocolumn,letterpaper,lengthcheck,superscriptaddress,showpacs,amssymb,amsmath,amsfonts,aps,altaffilletter,nofootinbib,nopreprintnumbers,longbibliography]{revtex4-1}

\usepackage{color}
\usepackage{hyperref}
\usepackage{amsmath}
\usepackage{amsthm}
\usepackage{multirow}
\usepackage{graphicx}
\usepackage{tikz}
\usetikzlibrary{calc}
\usepackage{placeins}
\usepackage{xspace}
\usepackage[normalem]{ulem}
\usepackage{physics,enumerate}

\usepackage{mathrsfs}
\DeclareSymbolFontAlphabet{\mathrsfs}{rsfs}

\newcommand{\lib}[1]{{\em #1\xspace}}

\newcommand{\MM}{\ensuremath{\mathcal{M}}} 
\newcommand{\DD}{\ensuremath{\mathcal{D}}} 
\newcommand{\RR}{\ensuremath{\mathcal{R}}} 

\newcommand{\Surf}{\ensuremath{\mathcal{S}}}
\def\insubscript{\rm inner}
\def\outsubscript{\rm outer}
\newcommand{\Sout}{\Surf_{\outsubscript}}   
\newcommand{\Sin}{\Surf_{\insubscript}}     
\newcommand{\Sone}{\Surf_{1}}               
\newcommand{\Stwo}{\Surf_{2}}               
\newcommand{\Sonetwo}{\Surf_{1,2}}
\newcommand{\Sthree}{\Surf_{2}^*}
\newcommand{\Sfour}{\Surf_{1}^*}

\newcommand{\Sthreei}{\Surf_{2}^{**}}
\newcommand{\Sfouri}{\Surf_{1}^{**}}
\newcommand{\Sfive}{\Surf_{0}^{*}}
\newcommand{\Sfivei}{\Surf_{0}^{**}}

\newcommand{\Sini}{\Surf_{\insubscript}^{*}}   
\newcommand{\Sinii}{\Surf_{\insubscript}^{**}} 
\newcommand{\SoneA}{\Surf_{1a}}             
\newcommand{\SoneB}{\Surf_{1b}}             
\newcommand{\SoneC}{\Surf_{1c}}             
\newcommand{\SoneD}{\Surf_{1d}}             
\newcommand{\StwoA}{\Surf_{2a}}             
\newcommand{\StwoD}{\Surf_{2d}}             
\newcommand{\SoneAB}{\Surf_{1a,b}}

\newcommand{\SoneABC}{\Surf_{1a,b,c}}
\newcommand{\StwoABC}{\Surf_{2a,b,c}}
\newcommand{\SoneABCD}{\Surf_{1a,b,c,d}}
\newcommand{\StwoABCD}{\Surf_{2a,b,c,d}}

\newcommand{\iSone}{\Surf_{1}^i}
\newcommand{\iStwo}{\Surf_{2}^i}

\newcommand{\HH}{\ensuremath{\mathcal{H}}}
\newcommand{\Hout}{\HH_{\outsubscript}}
\newcommand{\Hin}{\HH_{\insubscript}}
\newcommand{\Hone}{\HH_{1}}
\newcommand{\Htwo}{\HH_{2}}

\newcommand{\Hthree}{\HH_{2}^*}
\newcommand{\Hfour}{\HH_{1}^*}

\newcommand{\Hthreei}{\HH_{2}^{**}}
\newcommand{\Hfouri}{\HH_{1}^{**}}
\newcommand{\Hfive}{\HH_{0}^{*}}
\newcommand{\Hfivei}{\HH_{0}^{**}}

\newcommand{\Hini}{\HH_{\insubscript}^{*}}
\newcommand{\Hinii}{\HH_{\insubscript}^{**}}
\newcommand{\HoneA}{\HH_{1a}}
\newcommand{\HoneB}{\HH_{1b}}
\newcommand{\HoneC}{\HH_{1c}}
\newcommand{\HoneD}{\HH_{1d}}
\newcommand{\HtwoA}{\HH_{2a}}
\newcommand{\HtwoB}{\HH_{2b}}
\newcommand{\HtwoC}{\HH_{2c}}
\newcommand{\HtwoD}{\HH_{2d}}

\newcommand{\Nneg}{N_{-}}                   
\newcommand{\NnegO}{N_{-}^{0}}              

\def\tname{t} 
\def\ttouch{\tname_{\rm touch}}
\def\tbifurcate{\tname_{\rm bifurcate}}          
\def\tAH{\tname_{\rm bifurcate}^{\outsubscript}} 
\def\tSfour{\tname_{\rm bifurcate}^{\Sfour}}     
\def\tSfive{\tname_{\rm bifurcate}^{\Sfive}}     
\def\tStwo{\tname_{\rm annihilate}^{\Stwo}}      
\def\tcuspinii{\tname_{\rm cusp}^{\insubscript^{**}}} 

\def\ttouchval{3.86 M}    
\def\tAHval{0.702 M}      
\def\tSfourval{0.756 M}   
\def\tSthreeval{0.811 M}  
\def\tSfiveval{0.888 M}   
\def\tStwoval{5.35 M}     
\def\tcuspiniival{4.32 M} 
\def\tcuspfourival{4.9 M} 

\newcommand{\Gpp}{G_{\scriptscriptstyle{++}}}
\newcommand{\Gpm}{G_{\scriptscriptstyle{+-}}}

\newcommand{\Lslice}{L_\Sigma} 

\def\MOTSs{MOTSs\xspace}   
\def\MITSs{MITSs\xspace}   
\def\MOTTs{MOTTs\xspace}   

\begin{document}

\title[]{%
    Ultimate fate of apparent horizons during a binary black hole merger II:\\
    Horizons weaving back and forth in time
}

\author{Daniel Pook-Kolb}
\affiliation{
    Max-Planck-Institut f\"ur Gravitationsphysik (Albert Einstein Institute),
    Callinstr. 38, 30167 Hannover, Germany
}
\affiliation{
    Leibniz Universit\"at Hannover, 30167 Hannover, Germany
}

\author{Ivan Booth}
\affiliation{
    Department of Mathematics and Statistics, Memorial University of Newfoundland,
    St. John's, Newfoundland and Labrador, A1C 5S7, Canada
}

\author{Robie A. Hennigar}
\affiliation{
    Department of Mathematics and Statistics, Memorial University of Newfoundland,
    St. John's, Newfoundland and Labrador, A1C 5S7, Canada
}
\affiliation{
	Department of Physics and Astronomy, University of Waterloo, 
	Waterloo, Ontario, Canada, N2L 3G1
}
\affiliation{
	Department of Physics and Computer Science, Wilfrid Laurier University, 
	Waterloo, Ontario, Canada N2L 3C5
}


\begin{abstract}
    In this second part of a two-part paper, we discuss numerical
    simulations of a head-on merger of two non-spinning black holes.
    We resolve the fate of the original two apparent horizons by showing that
    after intersecting, their world tubes ``turn around'' and continue
    backwards in time.
    Using the method presented in the first paper \cite{PaperI} to locate these surfaces,
    we resolve several such world tubes evolving and connecting through
    various bifurcations and annihilations.
    This also draws a consistent picture of the full merger in terms of
    apparent horizons, or more generally, marginally outer trapped surfaces
    (\MOTSs).
    The MOTS stability operator provides a natural mechanism to 
    identify \MOTSs which should be thought of as black hole boundaries. 
    These are the two initial ones and the final remnant.
    All other \MOTSs lie in the interior and are neither stable nor inner
    trapped.

\end{abstract}

\maketitle

\section{Introduction}
\label{sec:intro}

The now numerous detections of gravitational wave events leave
little doubt that black hole coalescences are a regularly occurring
phenomenon in our universe
\cite{LIGOScientific:2018mvr,TheLIGOScientific:2016pea,
      Nitz:2018imz,Nitz:2019hdf,Nitz:2020naa,Venumadhav:2019lyq,
      Zackay:2019tzo}.
With the help of numerical relativity simulations,
the produced gravitational waves travelling to distant observers are
analysed and modeled with steadily increasing efficiency and accuracy
\cite{Pratten:2020ceb,Varma:2019csw}.
However, the details of the merger of the black holes
themselves is less well understood. 
This is partly for conceptual and partly for numerical reasons.

On the conceptual side, one needs to answer the question of how to describe
black holes in highly dynamical situations.
When a black hole is at rest or only slightly perturbed, the event horizon is
a suitable description \cite{Hawking:1972hy}.
In non-perturbative cases, however, its teleological nature makes
it unsuitable for gaining an understanding of the dynamics
\cite{Ashtekar:2004cn,Faraoni:2015pmn,Booth:2005qc,Hayward:2000ca}.
A much better alternative is provided by the quasilocal horizon framework
\cite{Ashtekar:2004cn,Krishnan:2007va,Ashtekar:2003hk}.
The central concept in this framework, the dynamical horizon, presents a
notion of black holes that is valid and satisfies physical laws even in the
highly nonlinear phases of the merger.
Dynamical horizons are based on the numerically accessible marginally outer
trapped surfaces (\MOTSs), i.e. surfaces $\Surf$ defined as having vanishing
outward expansion.
Following such a MOTS through the time evolution of a spacetime generates a
world tube, called a marginally outer trapped tube (MOTT).
In their original definition
\cite{Ashtekar:2004cn,Krishnan:2007va,Ashtekar:2003hk},
dynamical horizons are a certain subset of \MOTTs.
In the present work, MOTS stability \cite{Andersson:2005gq,Andersson:2007fh}
will play a central role and we shall refer to a stable MOTS as
{\em apparent horizon}\footnote{%
    The compatibility with the traditional terminology of apparent horizons as
    boundaries of trapped regions will be discussed below.
}
(AH) and to a MOTT foliated by apparent horizons as
{\em dynamical apparent horizon} (DAH).

Whether we consider dynamical horizons or \MOTTs in general,
one seemingly basic question remained open:
What happens to the horizons of two black holes when they merge?
It is well known that a common apparent horizon forms around the two
individual ones when they are sufficiently close to each other.
This common horizon immediately splits into an outer and an inner
branch.
This fact together with the observation that \MOTTs may in
principle weave back and forth in time (see e.g.
\cite{Hayward:2000ca,bendov}) sparked speculations that all the
horizons in a binary merger might, in fact, be parts of a single
world tube
\cite{Booth:2005ng,Moesta:2015sga,Gupta:2018znn}.

Prior to recent advances in methods for locating \MOTSs numerically
\cite{pook-kolb:2018igu}, it was not possible to further investigate these
ideas.
The world tubes that had to be tracked developed extremely distorted
shapes, which the typically used algorithms failed to resolve.
The fundamental assumption often employed to simplify the numerical
task is that the surface we wish to locate is star shaped,
i.e. it can be represented using an angle-dependent (coordinate)
distance function from some reference point.
See Ref.~\cite{Thornburg:2006zb} for a review.

By removing this limitation, it was shown recently that the
individual apparent horizons connect indirectly to the outer common
horizon by merging (non-smoothly) with the world tube of its inner
branch at the time when the individual apparent horizons touch
\cite{%
    PhysRevLett.123.171102,%
    PhysRevD.100.084044,%
    pook-kolb2020I,%
    pook-kolb2020II%
}.
However, they continue to exist afterwards and their later fate
was not fully resolved in these studies.
One reason for this is that in the utilized methods, one still had to
anticipate the possible shapes with appropriate initial guesses.
As we shall see, resolving their full fate requires even more exotic guesses,
and for this a new method was needed.
We have developed and presented such a method in the first paper of this
two-part series, henceforth denoted as paper I \cite{PaperI}.
As we shall see in the remainder of this  second paper, having such a
method is the key to resolving this question.

One might argue that in classical general relativity, whatever
happens in the interior of the event horizon remains -- by
definition -- causally disconnected from far away observers.
\MOTTs are always located in this interior region and,
after the outer common apparent horizon has formed, the individual apparent
horizons are even further away from it.
Nevertheless, the question of their fate seems highly relevant if
one aims at using \MOTTs to understand the merger.
To allow for a physically meaningful interpretation, these should
be well-behaved objects in the first place.
But how is this compatible with the results of paper I?
There, we have shown that a two-black-hole configuration may contain
a large number of \MOTSs not previously known.
Are these merely artifacts of an ``unphysical'' configuration?
Since they were found in time-symmetric initial data with a common AH already
present, neither the past nor the future of this configuration contains two
separate black holes.
In the present paper, we therefore look at the merger of initially
separate black holes and aim to answer two questions:
i) Do such additional \MOTSs also form dynamically in a merger of
initially separate black holes?
ii) What physical significance do they have and how can we
differentiate them from the intuitively more relevant apparent horizons we
associate with the individual black holes ($\Sonetwo$) and the final common
one ($\Sout$)?

To answer these questions, we perform numerical simulations
of the head-on collision of two non-spinning black holes with no
initial momentum.
In particular, we show explicitly that
i) such additional \MOTSs do form dynamically and
ii) the new \MOTTs we track do, in fact, bifurcate and annihilate
with the other \MOTTs.
This means that they are indeed weaving back and forth in time,
but as we will see later, they do not connect to form just one
single smooth surface.
A remarkable result is a surprisingly clear and predictable behavior
in terms of the stability of these \MOTTs:
Whenever a MOTT switches direction in time, it gains an
additional negative eigenvalue of the stability operator, i.e. it effectively
becomes ``more unstable''.
Furthermore, only three \MOTTs are stable (and hence DAHs) in the sense of
\cite{Andersson:2005gq,Andersson:2007fh} and are thus distinguished
from all other \MOTTs we find.
These are the two DAHs traced out by $\Sone$ and $\Stwo$ associated with the
individual black holes and the one final common DAH traced out by $\Sout$.

The rest of this paper is organized as follows.
We start by fixing the notation and introducing the required
mathematical concepts in Section~\ref{sec:basics}.
Section~\ref{sec:simulations} gives the numerical details of our
setup, the simulations and the method to locate and track the
marginal surfaces.
The new \MOTSs and the world tubes they trace out are introduced
in Section~\ref{sec:BLevolution}.
Here, we also describe a mechanism occurring multiple times along
otherwise smoothly evolving \MOTTs where a cusp forms followed by
a new self-intersection.
Additionally, we present \MOTSs of toroidal topology that exist inside the
individual \MOTSs $\Sone$ and $\Stwo$.
Section~\ref{sec:stabilityResults} connects the various observations
made in the previous sections with the MOTS stability properties.
The signature and the expansion of the ingoing null rays, together
important for understanding the behavior of the area, are presented
in Section~\ref{sec:signature}.
Finally, Section~\ref{sec:conclusions} will conclude with a
discussion of the main results.

\section{Basic Notions}
\label{sec:basics}

\subsection{Marginal surfaces and their world tubes}
\label{sub:MOTSs}

We consider four-dimensional spacetime
$(\MM, g_{\alpha\beta}, \nabla_\alpha)$ with Lorentzian four-metric
$g_{\alpha\beta}$ of signature $({-}\,{+}\,{+}\,{+})$.
For a smooth spacelike two-surface $(\Surf, q_{AB}, \DD_A)$, let
$\ell^\pm$ be two linearly independent future pointing null normals
scaled such that $\ell^+\cdot\ell^-=-1$.
In the present paper we will only consider closed surfaces $\Surf$
and we assume it is possible to assign an outward direction on
$\Surf$.
Then, $\ell^+$ is taken to be pointing outward and $\ell^-$ inward.
The expansions $\Theta_{\pm}$ of a congruence of null rays
travelling in the $\ell^\pm$ directions is then given by
\begin{equation}\label{eq:Theta}
    \Theta_{\pm} = q^{\alpha\beta} \nabla_\alpha \ell^\pm_\beta \,,
\end{equation}
where $q^{\alpha\beta} = e^\alpha_A e^\beta_B q^{AB}$ with
$e^\alpha_A$ being the pull-back from $\MM$ to $\Surf$.
The expansions can be seen as the trace of the extrinsic curvatures
$k^\pm_{AB}$ of $\Surf$ associated with $\ell^\pm$.
The (symmetric) trace-free part is given by the shear
\begin{equation}\label{eq:shear}
    \sigma^\pm_{AB}
        = \nabla_A \ell^\pm_B - \frac{1}{2}\Theta_\pm q_{AB}
        = e^\alpha_A e^\beta_B \nabla_\alpha \ell^\pm_\beta - \frac{1}{2}\Theta_\pm q_{AB}
        \,.
\end{equation}
We will call $\Theta_{+}$ the outgoing and $\Theta_{-}$ the ingoing
expansion.

The signs of $\Theta_{\pm}$ allow us to classify $\Surf$.
In particular, if $\Theta_{\pm} < 0$ then $\Surf$ is called a
{\em trapped surface}.
The existence of such a surface has been proven to imply that spacetime is
causally geodesically incomplete and thus singular
\cite{Penrose:1964wq}.
$\Surf$ is called a {\em marginally trapped surface} if $\Theta_{-}<0$
and $\Theta_{+} = 0$
and a {\em marginally outer trapped surface} (MOTS) if
$\Theta_{+}=0$ with no restriction on $\Theta_{-}$.
We mention here that
Andersson et al. show in~\cite{Andersson:2008up} that existence of a
strictly stable MOTS (introduced in Section~\ref{sub:stabilityTheory})
is sufficient for the singularity theorem mentioned above to hold.
Note that we can still scale the null normals by arbitrary positive
functions $f>0$ via
\begin{equation}\label{eq:nullrescaling}
    \ell^+ \to f\ell^+
    \quad\text{and}\quad
    \ell^- \to \frac{1}{f}\ell^- \;.
\end{equation}
Fortunately, the signs of the expansions $\Theta_{\pm}$ and
consequently the above characterization of $\Surf$ is invariant
under these transformations.

Let the  spacetime $\MM$ be foliated by spatial slices
$(\Sigma_t, h_{ij}, D_i, K_{ij})$ with Riemannian three-metric
$h_{ij}$ and extrinsic curvature $K_{ij}$.
Following a MOTS $\Surf$ through slices of the foliation provides
the notion of a MOTT.
More precisely, a smooth three-manifold $\HH$ is called a
{\em marginally outer trapped tube} (MOTT)
if it admits a foliation of \MOTSs.
Note that this definition of a MOTT makes no use of the foliation of spacetime
by the $\Sigma_t$.
We will, however, only consider \MOTTs $\HH$ with a foliation of \MOTSs
contained in the slices $\Sigma_t$.
For a spacelike future ($\Theta_{-}<0$) MOTT $\HH$, the foliation
can be shown to be unique \cite{Ashtekar:2005ez}.

The above objects are closely related to various terms involving the word
``horizon''.
For instance, a spacelike future MOTT is defined as a
dynamical horizon in \cite{Ashtekar:2003hk,Ashtekar:2002ag}
while in \cite{pook-kolb2020I,pook-kolb2020II}, the definition of a
dynamical horizon has been generalized to refer to any MOTT.
Additional qualifiers (e.g. future, spacelike) were used to then specialize
where needed.
One of the reasons for this generalization is that \MOTSs and \MOTTs were
found to appear in a much wider variety than expected and the
original dynamical horizon did not cover all the interesting cases.
However, the examples in \cite{pook-kolb2020I,pook-kolb2020II} were just the
start and \cite{Booth:2020qhb} showed that even in a single slice of the
Schwarzschild spacetime there are an infinite number of \MOTSs.
This together with the results shown in the first and this second paper
clearly suggest that not all of these objects should be thought of as black
hole boundaries.
As mentioned in the introduction, the notion of MOTS stability turns
out to reliably select those \MOTSs which possess reasonable physical
properties and can thus be called horizons.
We will therefore introduce the MOTS stability operator in the following
section.

\subsection{MOTS stability}
\label{sub:stabilityTheory}

The concept of MOTS stability in the sense of
Refs.~\cite{Andersson:2005gq,Andersson:2007fh} is helpful to get a
deeper insight into the evolution properties of a MOTS.
In particular, it will be useful in assessing at which time $t$ a
MOTT $\HH$ becomes tangent to a spatial slice $\Sigma_t$ and
thus generically ``turns around'' in time.
These instances correspond to a MOTT appearing and bifurcating
into two branches, as well as when two \MOTTs merge and
annihilate.
Furthermore, we expect the physically relevant horizons to be
boundaries for trapped and untrapped surfaces, at least in a
neighbourhood. This property also turns out to be closely related to
the notion of MOTS stability.

Before delving into the fully generic case, it will be helpful to first
consider a very simple situation in which the stability operator appears
almost naturally.
Spherically symmetric spacetimes provide such an example and so here we
restrict our attention to spherically symmetric \MOTSs.

\subsubsection{MOTTs in spherical symmetry}

\begin{figure}
    \includegraphics{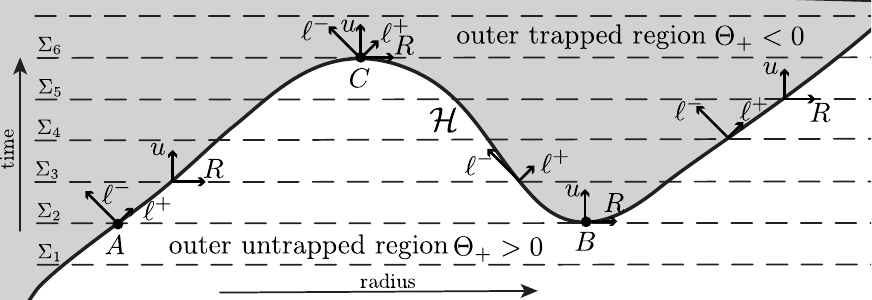}
    \caption{\label{MTTjump}%
        Spherically symmetric evolution of a MOTT $\HH$ (thin solid line).
        It is perfectly possible for $\HH$ to weave back and forth through
        the time foliation. See, for example, \cite{Booth:2005ng,bendov}
        for exact solutions exhibiting this behaviour.
    }
\end{figure}

We will start by establishing some basic properties of \MOTTs
in spherical symmetry.
In this setting, each point in the two-dimensional $(t,r)$ space represents a
sphere and we can calculate its expansion $\Theta_+$.
A point $(t,r)$ can then be labeled as outer trapped, outer untrapped or
marginally outer trapped as shown in FIG.~\ref{MTTjump}.
A MOTT $\HH$ traces a curve through the $(t,r)$ space and it can weave its way
back and forth through the foliation of spacetime.

Tangents to this curve can be written in the form
\begin{align}\label{cV}
    \mathcal{V}^\alpha = N (u^\alpha + v R^\alpha) \;,
\end{align}
where $u^\alpha$ and $R^\alpha$ are respectively the unit vectors in the $t$
and $r$ directions. $N$ can be any function of $(t,r)$, though a convenient
choice is the lapse. $v$ is the speed of $\HH$ relative to the foliation.
$\HH$ is spacelike if $|v|>1$, timelike if $|v|<1$ and null if $|v| = 1$. If
$\HH$ becomes tangent to the foliation (as at $B$ or $C$) then at that point
$|v| \rightarrow \infty$.

The speed $v$ may also be calculated from the fact that $\Theta_+ = 0$ on
$\HH$. Then writing the derivative in the $\mathcal{V}$ direction as
$\delta_{\mathcal{V}}$ (the notation is chosen to be compatible with the next
section), $\delta_{\mathcal{V}} \Theta_+ = 0$. It follows that
\begin{align}\label{vdet}
    v = -\frac{\delta_u \Theta_+}{\delta_R \Theta_+}
      = 1 - 2 \frac{\delta_{\ell^+} \Theta_+}{\delta_R \Theta_+}
    \;,
\end{align}
where we made a particular choice for the null vectors
relative to $u$ and $R$:
\begin{align}\label{eq:nullchoice}
    \ell^+ = \frac{1}{2} (u + R) \mbox{ and }  \ell^-= u - R \;.
\end{align}
Then we can again see that if $\delta_R \Theta_+ \rightarrow 0$,
$|v| \rightarrow \infty$ (as long as $\delta_{\ell^+} \Theta_+ \neq 0$).

Next, applying the spherically symmetric null Raychaudhuri equation we find
\begin{align}\label{v2}
    v = 1 + 2 \frac{\Gpp}{\delta_R \Theta_+} \;.
\end{align}
By the null energy condition,
$\Gpp = G_{\alpha \beta} \ell_+^{\alpha} \ell_+^{\beta} \geq 0$.
If it vanishes, then $v=1$ and so $\HH$ is outward null and isolated at that
point. If matter falls through $\HH$
($G_{\alpha \beta} \ell_+^{\alpha} \ell_+^{\beta}>0$)
and the region outside of $\HH$ is outer untrapped
(from $A$ to $C$ and then $B$ onwards in FIG.~\ref{MTTjump})
then $\delta_R \Theta_+ > 0 \; \Rightarrow \; v>1$ and $\HH$ is spacelike
outward at that point.
However if the region outside of $\HH$ is outer trapped (from $C$ to $B$) then
$\delta_R \Theta_+ < 0$ and so $\HH$ could be either spacelike, timelike or
(inward) null.
At $C$ it transitions from $v=\infty$ to $-\infty$ and at $B$ from
$-\infty$ to $\infty$.

These quantities also determine the expansion of $\HH$. From \eqref{cV} and \eqref{v2}
\begin{align}
    \Theta_{\mathcal{V}}
        = N \left( \Theta_u  + v \Theta_R \right)
        = - \left( \frac{\Gpp}{\delta_R \Theta_+} \right) \Theta_- \;.
\end{align}
Thus if $\Gpp=0$, then $\Theta_{\mathcal{V}}$ is
non-expanding. However if $\Gpp \neq 0$ and $\Theta_- < 0$
(that is, the $(t,r)$ spheres get smaller moving inward from the MOTT),
then $\HH$ is expanding if the region just outside is outer untrapped (from
$A$ to $C$ and then $B$ onwards) and shrinking if the region outside of $\HH$ is
outer trapped (from $C$ to $B$). Together these results mean that if we
consider $\HH$ as a continuous curve running in the direction $A$ to $B$ then
it is increasing in area as one would expect for a black hole horizon.

Note that the exact location of the transition points is at least partly a
function of the foliation. For example, if we chose a foliation rotated relative to
our original $\Sigma_t$ in FIG.~\ref{MTTjump}, then the points of tangency
between $\HH$ and the foliation would be different. Their number could even
increase. However the possible timelike signature of $\HH$ means that this
wending through time cannot (always) be understood as simply a by-product of
the choice of spacetime foliation. Timelike sections necessarily intersect
with many slices of any foliation.

Finally note that a simulation that only tracked outermost \MOTSs would see an
apparent horizon jump from $A$ to $B$ at $\Sigma_2$. However a more careful
tracker would identify $B$ as a MOTS pair creation event with one surface subsequently
moving inwards while the other expands outwards. Ultimately that inward moving
MOTS would annihilate with the original apparent horizon at $C$. Both pair
creation and annihilation events happen at points where
$\delta_R \Theta_+ \rightarrow 0$.

\subsubsection{The MOTS stability operator}
\label{sub:stabilityGeneral}

Away from spherical symmetry everything is more complicated, though many of
the themes (and conclusions) that we have just examined remain. In particular
$\delta_{R} \Theta_+$ continues to play a key role though it is now
generalized to become the MOTS
stability operator \cite{Andersson:2005gq,Andersson:2007fh}.

Leaving rigid spherical symmetry behind, there is no longer just one way to
expand a surface outwards (or inwards).
Since deformations tangent to a MOTS $\Surf$ leave $\Theta_{+}$ invariant, we
will only consider deformations along some direction $V^\alpha$ normal to
$\Surf$. In principle, this need not be limited to a particular slice
$\Sigma_t$ containing $\Surf$.
Now, for such a deformation of $\Surf$,
consider a family $\Surf_\nu$ of similar surfaces such that $\Surf_0 = \Surf$.
Then we can consider the vector field of normal vectors $V^\alpha$ to these
surfaces.
In turn these generate a congruence of curves that map points between the
$\Surf_\nu$ and we can write the tangent vector to these curves, the generator
of the deformations, as
\begin{equation}\label{eq:partialnu}
    \frac{\partial}{\partial\nu} = \psi V
\end{equation}
for a function $\psi$.
Note that the local deformation of $\Surf$ is fully determined by $\psi$ as a
function on just $\Surf$.
For the spherical deformations of the last section, $\psi$ would be a
constant and $V^\alpha$ would be the normal $R^i$ in a $t=\text{const}$ slice.

Next, to each of the $\Surf_\nu$ we can construct a (non-unique) pair of null
normals $\ell^\pm_\nu$ and calculate the expansions $\Theta_{\pm}^\nu$.
They are chosen so that $\ell^\pm_0 = \ell^\pm$ (the original null normals on
$\Surf$). The MOTS stability operator $L_V$ with respect to the normal
$V^\alpha$, scaled\footnote{%
    In particular, $V$ cannot be parallel to $\ell^+$.
}
such that $V^\alpha \ell^+_\alpha = 1$,
is then defined as the derivative of $\Theta_+^\nu$ with respect to
$\nu$:
\begin{equation}\label{eq:stabilityDef}
    L_V \psi := \delta_{\psi V} \Theta_{+}
        := \frac{\partial}{\partial\nu}\Big|_{\nu=0} \Theta_{+}^\nu \;.
\end{equation}
It is shown in \cite{Andersson:2007fh} that $L_V$ does not
depend on the choice of $\ell^\pm_\nu$ away from $\Surf$.
On the other hand, the definition of $L_V$ is 
not invariant under
the rescalings~\eqref{eq:nullrescaling} of the null normals $\ell^\pm$.
However, since we will be interested only in its eigenvalues, we can use
the fact that $L_V$ is isospectral under \eqref{eq:nullrescaling}
as shown in \cite{Jaramillo_2015} and we shall work with the particular choice
\eqref{eq:nullchoice}.

Just as in the spherical case of the previous section, we will
discuss the evolution properties of a MOTT $\HH$ in the context
of some fixed foliation $\Sigma_t$ and use this foliation to talk about
{\em bifurcations} and {\em annihilations}.
We will then need to choose the vector $V^\alpha$ as the normal of $\Surf$
which lies in $\Sigma_t$, i.e. we choose\footnote{
    The factor $2$ results from our different convention \eqref{eq:nullchoice}
    for the scaling of $\ell^+$ as compared to \cite{Andersson:2005gq}.
    However, this does not change the spectrum of $L_V$.
}
$V^\alpha = 2 R^\alpha$.
The stability operator with respect to the slice $\Sigma_t$ is then defined as
$\Lslice := L_{2 R^\alpha}$ and it takes the form
(e.g. \cite{Newman_1987,Andersson:2005gq,Andersson:2007fh}):
\begin{align}\label{eq:stab}
    \Lslice\psi = -\tilde{\triangle} \psi + \left(
        \frac{1}{2}\RR - 2|\sigma_+|^2 - 2\Gpp - \Gpm
    \right) \psi \,,
\end{align}
where factors of $2$ differing from \cite{Andersson:2005gq,Andersson:2007fh}
result from our different cross normalization $\ell^+\cdot\ell^-=-1$ and
where
\begin{align}\label{eq:triangle}
    \tilde{\triangle} \psi = (\mathcal{D}_A - \omega_A)(\mathcal{D}^A - \omega^A) \psi
\end{align}
with $\omega_A = -e_A^\alpha \ell^-_\beta \nabla_\alpha \ell_+^\beta$ the connection
on the normal bundle of $\Surf$, $\mathcal{R}$ its Ricci scalar,
$\Gpm = G_{\alpha \beta} \ell_+^\alpha \ell_-^\beta$, $\Gpp = G_{\alpha \beta} \ell_+^\alpha \ell_+^\beta$
and
$|\sigma_+|^2 = \sigma^+_{AB} \sigma^{AB}_+$.
This is a second order, linear, elliptic operator with discrete spectrum and
for a non-vanishing connection $\omega_A$ it is not self-adjoint.
However, its principal eigenvalue $\lambda_0$, i.e. the eigenvalue with the
smallest real part, is always real.
A MOTS $\Surf$ with $\lambda_0 \geq 0$, $\lambda_0 > 0$, $\lambda_0 < 0$
is called {\em stable}, {\em strictly stable}, or {\em unstable}, respectively.
The meaning of this terminology will soon become clear.

Note that for the spherical cases that we considered in the last section,
$\omega_A = 0$ and $\sigma^+_{AB} = 0$ and we were only considering constant
$\psi$. With those restrictions
\begin{align}\label{eq:LsigmaSpherical}
    \Lslice \psi
        = 2\,\delta_{\psi R} \Theta_+
        = \left( \frac{1}{R^2} - 2\Gpp - \Gpm \right) \psi \;,
\end{align}
where $R$ is the areal radius of the MOTS. Then the term in parentheses is
the principal\footnote{%
    Allowing non-constant $\psi$ would add a term $-\Delta_\Surf\psi$, where
    $\Delta_\Surf$ is the Laplacian on $\Surf$. The spectrum would then be
    that of the Laplacian on a round sphere (with lowest eigenvalue being zero)
    shifted by the constant term in parenthesis in Eq.~\eqref{eq:LsigmaSpherical}.
}
eigenvalue of $\Lslice$.
Hence in spherical symmetry we can only have an ``ingoing'' $\HH$ when the sum
of the matter terms is comparable in size to $1/R^2$.
If it is not then, as we saw earlier, $\HH$ is necessarily spacelike and (if
$\Theta_- < 0$) expanding.
However if it is equal so that $\delta_{\psi R} \Theta_+ = 0$, then we can
have horizon pair formation (annihilation) as at $B$ ($C$) in
FIG.~\ref{MTTjump}.
Physically this can be thought of as a case where the matter outside is dense
enough to cause the formation of a new horizon outside the old
one \cite{Booth:2005ng}.

\begin{figure}
    \scalebox{0.63}{\includegraphics{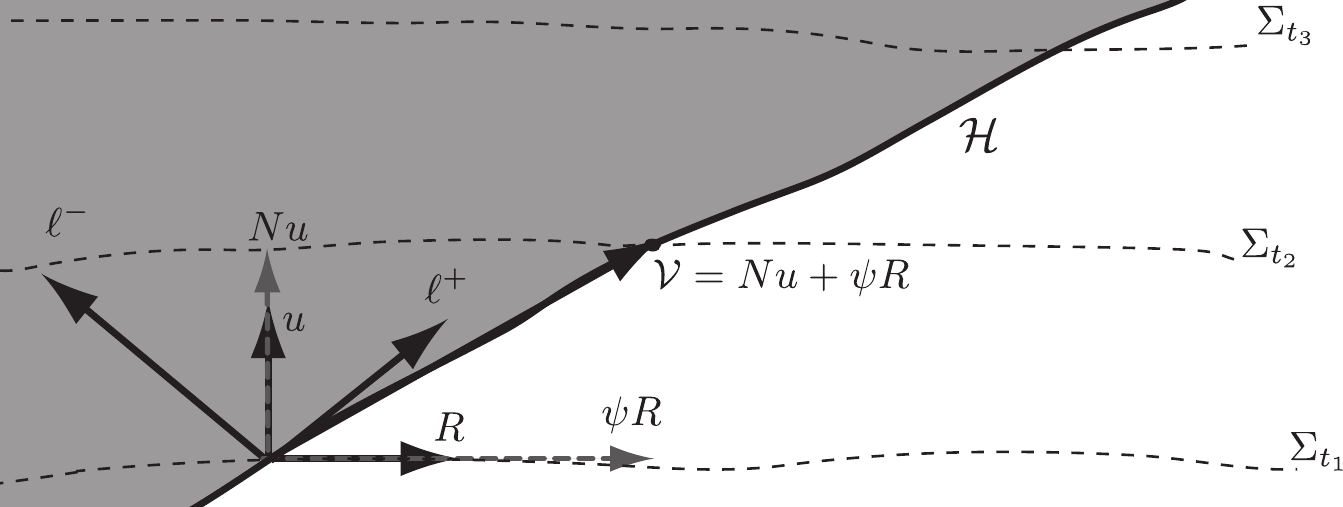}}
    \caption{\label{fig:evolve}%
        Definition of the evolution vector $\mathcal{V}$ along a MOTT $\HH$.
        $\mathcal{V}$ can be split into a component $\psi R$ orthogonal to
        $\Surf = \HH \cap \Sigma_{t_1}$ within $\Sigma_{t_1}$ and one
        orthogonal to $\Sigma_{t_1}$, i.e. $Nu$, which is also orthogonal to
        $\Surf$.
    }
\end{figure}

Returning to the general stability operator (\ref{eq:stab}), Andersson et al.
proved in \cite{Andersson:2005gq, Andersson:2007fh} that $\lambda_0 > 0$
implies existence of a smooth MOTT $\HH$ containing $\Surf$.
That is, $\Surf$ evolves smoothly to the future and to the past for at least a
short time interval.
This can be understood intuitively in the following way.
Consider a three-dimensional MOTT $\HH$ in spacetime, illustrated in
FIG.~\ref{fig:evolve}.
Again, let $\mathcal{V}^\alpha$ be the tangent to $\HH$
orthogonal to the MOTSs $\Surf$ and scaled such that
$\mathcal{L}_\mathcal{V} t = 1$, where $\mathcal{L}_\mathcal{V}$ is the
Lie-derivative along $\mathcal{V}^\alpha$.
For a fixed foliation $\Sigma_t$,
we can split $\mathcal{V}^\alpha$ into the components orthogonal and tangent to
$\Sigma_t$,
\begin{equation}\label{eq:V3plus1}
    \mathcal{V}^\alpha = N u^\alpha + \psi R^\alpha \;,
\end{equation}
where the foliation fully determines the lapse $N$ and so $Nu^\alpha$ is fixed.
%
Clearly, the variation of the expansion along the MOTT $\HH$ vanishes,
\begin{equation}\label{eq:vary_expansion_along_H}
    \delta_{\mathcal{V}} \Theta_+ = 0 \;,
\end{equation}
which means that
finding the tangent vector $\mathcal{V}^\alpha$ amounts to solving the
inhomogeneous partial differential equation
\begin{align}\label{eq:evol}
    \Lslice \psi = -2\delta_{N u} \Theta_{+}
\end{align}
for $\psi$.
The inhomogeneity on the righthand side of this is fixed by the foliation.

Note that if $\Lslice$ is invertible, then there exists a solution for
\eqref{eq:evol} for {\em any} lapse $N$.
This is guaranteed if, e.g., $\lambda_0>0$.
However, if $\lambda_0 \leq 0$, then invertibility can fail, as eigenvalues may
vanish.
Equivalently, in that case there are homogeneous solutions to the evolution
equation \eqref{eq:evol}:
that is we can choose a $\mathcal{V}^\alpha$ that is tangent to $\Sigma_t$ and still
satisfies \eqref{eq:vary_expansion_along_H}. 
The pair creation event $B$ in FIG.~\ref{MTTjump} is such an event. 
This association of vanishing eigenvalues with pair creation/annihilation has
also been observed away from spherical symmetry \cite{pook-kolb:2018igu}.
In the present paper, we shall see multiple instances of horizons appearing
and vanishing precisely as one of the eigenvalues -- not necessarily
$\lambda_0$ -- becomes zero.
If it is $\lambda_0$ that vanishes, then Proposition~5.1 of \cite{Andersson:2008up}
shows that generically $\HH$ is tangent to $\Sigma_t$ and, for fixed slicing,
unique at least in a neighbourhood.


Note too that $\lambda_0>0$ implies that $\delta_{\psi R} \Theta_+>0$ and
again, just as we saw in the spherically symmetric case, this is sufficient to
imply
that the evolving $\HH$ is spacelike at that point
\cite{Andersson:2005gq,Andersson:2007fh} and so, if $\Theta_-<0$, expanding
(see also \cite{Hayward:1993wb,Ashtekar:2002ag}). This area expansion theorem
has been generalized to include timelike ``backwards in time'' segments
\cite{Bousso:2015mqa}. However this is not the end of the story: certain
assumptions made in that generalized proof have now been shown to not always
hold during black hole mergers \cite{pook-kolb2020I,pook-kolb2020II}.
We will see further examples of this in Section \ref{sec:signature}.

%
%
%
%

Stability is not only useful for understanding the evolution of $\HH$, it
also tells us something about local properties of $\Surf$ {\em within} the
slice $\Sigma_t$.
In paper I, it  is shown that the stability operator for MOTS can be understood as the 
analogue of the Jacobi operator for geodesics \cite{PaperI}. Thus if one considers an axisymmetric MOTS to be one
of a congruence of marginally outer trapped (possibly open) surfaces, then the number of 
negative eigenvalues of the stability operator corresponds to the number of intersections with other nearby members of
the congruence. Hence a stable MOTS with $\lambda_0>0$ does not intersect its neighbours while an unstable one certainly
does have such intersections.

More generally, it is shown in \cite{Andersson:2005gq,Andersson:2007fh}
that a strictly stable MOTS $\Surf$ ($\lambda_0>0$) has the {\em barrier property}.
In essence, this means that given a close-by surface $\Surf'$
then if $\Surf'$ has expansion $\Theta'_{+}\leq0$, it cannot extend
into the exterior of $\Surf$. Similarly if $\Theta'_{+}\geq0$, it
cannot enter the interior.
On the other hand if $\Surf$ has the barrier property, it is at least stable.

Following from these results, we adopt the following convention in the present work. 
A stable MOTS $\Surf$ shall be called an {\em apparent horizon}
(AH).
Note that an apparent horizon is usually defined as the outer boundary of the
trapped region in a given slice $\Sigma_t$.
Due to Theorem~2.1 in \cite{Andersson:2008up}, this boundary is a stable
MOTS, so our definition includes the previous one
and extends it to include surfaces that can still reasonably be associated
with black hole boundaries. Examples are the previously outermost \MOTSs
$\Sone$ and $\Stwo$, as we shall see below.
Similarly, a three-surface $\HH$ will be called a
{\em dynamical apparent horizon} (DAH) if it allows a foliation of
apparent horizons.

\subsubsection{Simplification for vacuum, axial symmetry and no spin}
\label{sec:stabilityAxisym}

As shown in the first paper, for non-spinning
axisymmetric \MOTSs $\Surf$ in vacuum, we can simplify $\Lslice$ to
\begin{equation}\label{eq:stabilityAxisym}
    \Lslice\psi = -\Delta_\Surf \psi + \left(
        \frac{1}{2}\RR - 2|\sigma_+|^2
    \right) \psi
    \,,
\end{equation}
where $\Delta_\Surf=\DD_A\DD^A$ is the Laplacian on $\Surf$.
In this case $\Lslice$ is self-adjoint with purely real spectrum.
Another simplification can be made if we introduce coordinates
$(\theta,\phi)$ on $\Surf$. Let $\phi$ be the coordinate along
orbits of the axial Killing field $\varphi^A$ preserving the
two-metric $q_{AB}$ and vanishing at exactly two points, the poles
of $\Surf$. We take $\phi$ to be in the range $[0,2\pi)$.
Let further $\theta$ be any coordinate orthogonal to $\phi$, e.g.
$\cos\theta=\zeta$, with $\zeta$ constructed as in
Ref.~\cite{Ashtekar:2004gp}.
Then we can write any eigenfunction $\psi$ of $\Lslice$ as
\begin{equation}\label{eq:psifourier}
    \psi(\theta,\phi) = \sum_{m=-\infty}^{\infty} \psi_m(\theta) e^{im\phi}
    \,.
\end{equation}
For each $m\in\mathbb{Z}$, the eigenvalue problem
$\Lslice\psi=\lambda\psi$ then reduces to a one-dimensional
problem
\begin{equation}\label{eq:stability1D}
    \Lslice^m \psi_m := (\Lslice + m^2 q^{\phi\phi}) \psi_m = \lambda \psi_m
    \,.
\end{equation}
The eigenvalues of Eq.~\eqref{eq:stability1D} are labeled as
$\lambda_{l,m}$, where $l$ is chosen to run from $l=|m|$ over the
eigenvalues in ascending order.
This guarantees that for a round sphere, for which the spectrum reduces
to that of the Laplacian on a sphere of radius $R_0$ and shifted by
$\frac{1}{2}\RR=1/R_0^2$, the eigenvalues are labeled in
the conventional way as $\lambda_{l,m} = (1+l(l+1))/R_0^2$.
For brevity, we will sometimes write $\lambda_l := \lambda_{l,0}$.

\section{Numerical setup and MOTS finding}
\label{sec:simulations}

We use Brill-Lindquist initial data \cite{Brill:1963yv} for our
simulations. These describe a Cauchy slice $\Sigma$ which is time
symmetric, i.e. with vanishing extrinsic curvature.
The topology of $\Sigma$ is
$\mathbb{R}^3\setminus\left\{x_1,x_2\right\}$, where $x_{1,2}$ are the
coordinates of two punctures.
The Riemannian three-metric is conformally flat,
$h_{ij} = \psi^4\delta_{ij}$, where $\delta_{ij}$ is the flat metric.
The conformal factor is given by
\begin{equation}\label{eq:BLmetric}
    \psi = 1 + \frac{m_1}{2r_1} + \frac{m_2}{2r_2} \,,
\end{equation}
where $m_{1,2}$ are the bare masses of the black holes and $r_{1,2}$
are the (coordinate) distances to the respective puncture.
We shall here focus primarily on one particular configuration with
total ADM mass $M = m_1 + m_2 = 1$ and a mass ratio of
$q = m_2/m_1 = 2$ (i.e. $m_1=1/3$, $m_2=2/3$).
We choose a distance parameter of $d:=\lVert x_2 - x_1\rVert_2=0.9$
resulting in two black holes which are initially separate with no
common apparent horizon present.

We track the various \MOTSs in the simulations using the method
described in
\cite{pook-kolb:2018igu,PhysRevD.100.084044} and available from
\cite{pook_kolb_daniel_2021_4687700},
which in turn uses software libraries described in
\cite{erik_schnetter_2019_3258858, mike_boyle_2018_1221354, 2020SciPy-NMeth,
      van_der_Walt_NumPy, mpmath, meurer2017sympy, Hunter:2007,
      michael_droettboom_2018_1202077}.
Two approaches are used to locate the new \MOTSs in the simulations.
One is the shooting method described in paper I \cite{PaperI}, which can,
in principle, locate all axisymmetric \MOTSs of spherical topology by choosing
suitable starting points on the $z$-axis.
This method has been implemented in \cite{pook_kolb_daniel_2021_4687700} and
can be applied to both, analytically known initial data as well as to slices
obtained e.g. from numerical simulations.
The other method for generating initial guesses is motivated by the assumption
that \MOTSs may vanish and appear only in pairs of two.
Based on this idea, we try to track each MOTS to the future and to the past.
Whenever a MOTS cannot be tracked further in either direction, we look for a
``close by'' one with which it might merge.
Such a merger will be an annihilation if it happens to the future
and a bifurcation if it happens in the past direction.
Appendix~\ref{app:CESs} details the method we use to locate such a corresponding
MOTT using families of surfaces of constant expansion $\Theta_{+}$.

The simulations themselves are carried out using the
\lib{Einstein Toolkit} \cite{Loffler:2011ay,EinsteinToolkit:web}.
The Brill-Lindquist initial data are generated by
\lib{TwoPunctures} \cite{Ansorg:2004ds},
while we use an axisymmetric version of
\lib{McLachlan} \cite{Brown:2008sb}
for evolving these data in the BSSN formulation of the
Einstein equations.
This uses \lib{Kranc} \cite{Husa:2004ip,Kranc:web}
to generate efficient C++ code.
We always work with the
$1+\log$ slicing and a $\Gamma$-driver shift condition
\cite{Alcubierre:2000xu,Alcubierre:2002kk}.
An important feature of these gauge conditions is that they are ``singularity
avoiding'', which results in simulation time effectively slowing down close to
the punctures.
It was seen in \cite{evans:2020lbq} that the individual \MOTTs $\Sonetwo$
essentially stop evolving due to this effect, with the precise behavior being
highly dependent on the choice of initial data:
A smaller initial distance allows the $\Sonetwo$ to evolve further during the
simulation.
We repeat our simulations at different resolutions to ensure
convergence of our results.
Most of these results are obtained using a spatial grid resolution
of $1/\Delta x = 720$, with additional simulations carried out with
$1/\Delta x = 240$, $360$ and $480$. Shorter simulations to verify
certain features were performed at
$1/\Delta x = 960$ and $1920$.
We do not use mesh refinement and choose our domain large enough to
ensure that any boundary effects do not reach the \MOTSs for as
long as we track them.
Additional details and convergence properties of our simulations are described
in Ref.~\cite{PhysRevD.100.084044}.

\section{Overview of the various \MOTTs}
\label{sec:BLevolution}

The general picture one expects to find in head-on mergers of two
black holes has previously been analyzed in great detail
\cite{Mourier:2020mwa,PhysRevLett.123.171102,PhysRevD.100.084044,pook-kolb2020I,pook-kolb2020II}:
Initially, only two individual apparent horizons are present, $\Sone$ and $\Stwo$,
belonging to the two separate black holes.
At a time $\tAH$, common \MOTSs $\Sout$ and $\Sin$ form as one surface and
bifurcate into two branches.
While $\Sout$ settles to the final Schwarzschild horizon, $\Sin$ travels
inwards and becomes increasingly distorted.
At the precise time when $\Sonetwo$ touch, denoted as $\ttouch$, $\Sin$
forms a cusp and coincides with $\Sone \cup \Stwo$.
Immediately afterwards, $\Sonetwo$ intersect each other while $\Sin$ forms
self-intersections.
However, the final fate of $\Sone$, $\Stwo$ and $\Sin$ had not been resolved in those
studies. We shall attempt to resolve that fate here.

In the following, we will encounter several new \MOTSs $\Surf$ and we will, as
before, differentiate between them using different sub- and superscripts.
It is understood that replacing ``$\Surf$'' with ``$\HH$'' indicates that we
refer to the MOTT traced out by $\Surf$.

\subsection{Area evolution}
\label{sub:bifurcations}

\begin{figure}
    \includegraphics[width=\linewidth]{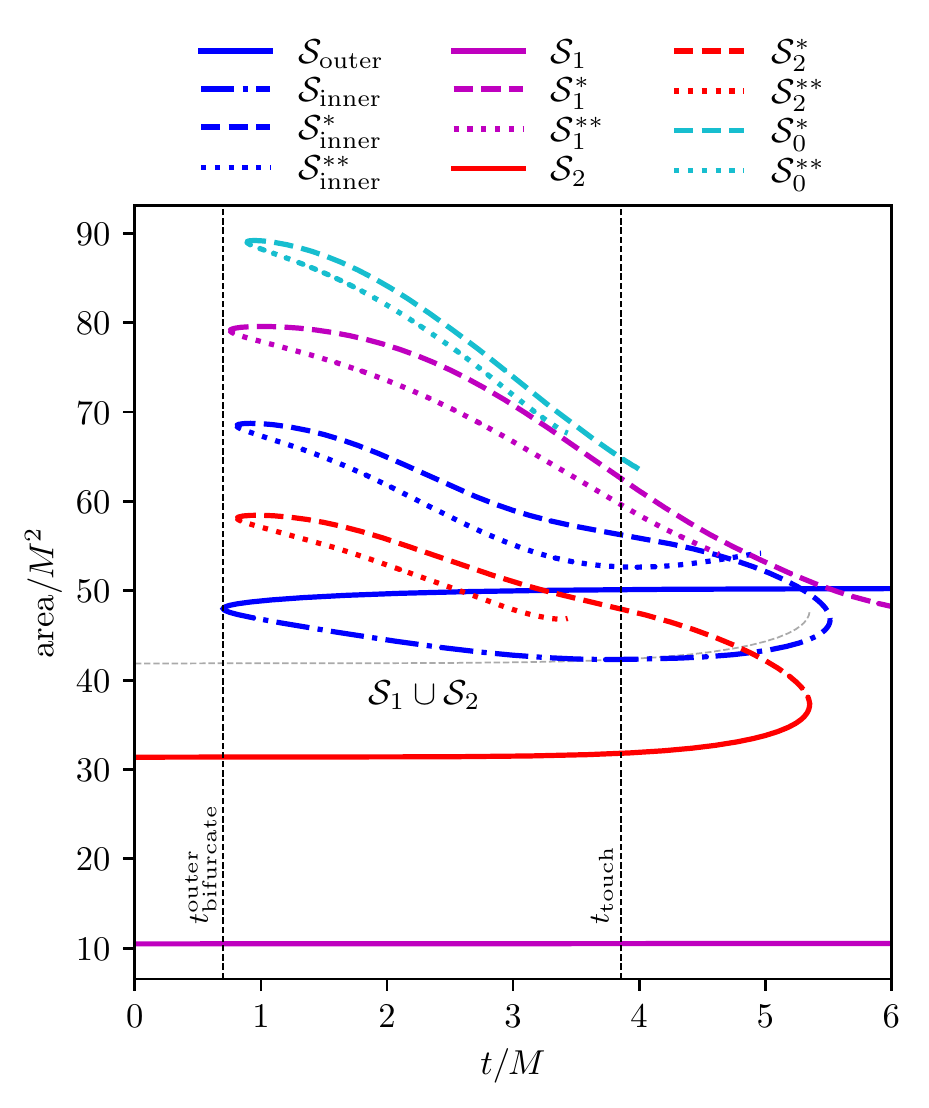}%
    \caption{\label{fig:BL31_areas_overview}%
        Evolution of the area of the various \MOTSs.
        Lines of the same color and different line styles smoothly connect and
        correspond to a single world tube continuing back and forth in time
        (except for the pair $(\Sone, \Sfour)$, which is discussed in
        Section~\ref{sub:worldtube_S1}).
        The sum of the areas of $\Sonetwo$ is shown as the thin dashed line.
        This coincides with the area of $\Sin$ at $\ttouch$ marking the merger
        of $\Sin$ with $\Sone \cup \Stwo$.
        Note that despite some \MOTSs having larger area than $\Sout$, they
        are, in fact, all contained within $\Sout$ for $t\geq\tAH$.
        For all curves that end without smoothly connecting to another curve, we lost
        track of the corresponding MOTS for numerical reasons
        (see the end of Section~\ref{sub:bifurcations} for details).
    }
\end{figure}

The main results are most easily visualized in terms of the area of
the various \MOTSs.
FIG.~\ref{fig:BL31_areas_overview} shows that we indeed find multiple new
\MOTSs previously not known, each forming in a bifurcation as a pair with an
{\em outer} and an {\em inner} branch.
Furthermore, we find \MOTSs which merge and annihilate in pairs of two.
Each bifurcation and annihilation connects two \MOTTs in a locally smooth
world tube.
Upon formation, the area of the outer branch increases while it decreases for
the inner branch.
However, with the exception of $\Sout$, even the areas of the outer branches
soon start to decrease.
This non-monotonic behavior of the area along a smooth portion of a MOTT has
been previously discussed for $\Sin$ in
\cite{PhysRevLett.123.171102, pook-kolb2020I} and has been attributed to
properties of the expansion $\Theta_{-}$ of the ingoing null rays $\ell^-$ and
to the signature of the world tube.
We shall here extend the discussion of these properties to the new world tubes
in Section~\ref{sec:signature}.
One characteristic that all \MOTSs $\Surf$ along such a MOTT have in common
is which punctures they enclose, i.e. no MOTS crosses a puncture in
its evolution.
Further, all additional bifurcations happen {\em after} the formation of the outermost
common MOTS $\Sout$ at $\tAH$ and the new \MOTSs are solely contained within
$\Sout$.

We note here that we lose track of  some of the \MOTSs, such as $\Sthreei$, during the
simulation without having an indication of an annihilation.
The reason we cannot track these \MOTSs further is a purely numerical one.
The shapes move very close to one of the punctures in our numerical
coordinates (the proper distance to the puncture is, of course, always infinite).
This close proximity results in loss of numerical accuracy since a large
spatial region is covered by a decreasing number of numerical grid points as
the puncture is approached, i.e. this region is numerically underresolved.
Fortunately, the analysis of the MOTS stability spectrum enables us to clearly
differentiate between a MOTS vanishing due to annihilation and one vanishing
due to loss of accuracy.

\subsection{The world tubes}
\label{sub:worldtubes}

In this subsection, we will give an overall description of the individual
connected world tubes.

\subsubsection{The world tube of $\Sout$}
\label{sub:worldtube_Sout}

\begin{figure*}
    \includegraphics[width=0.333\linewidth]{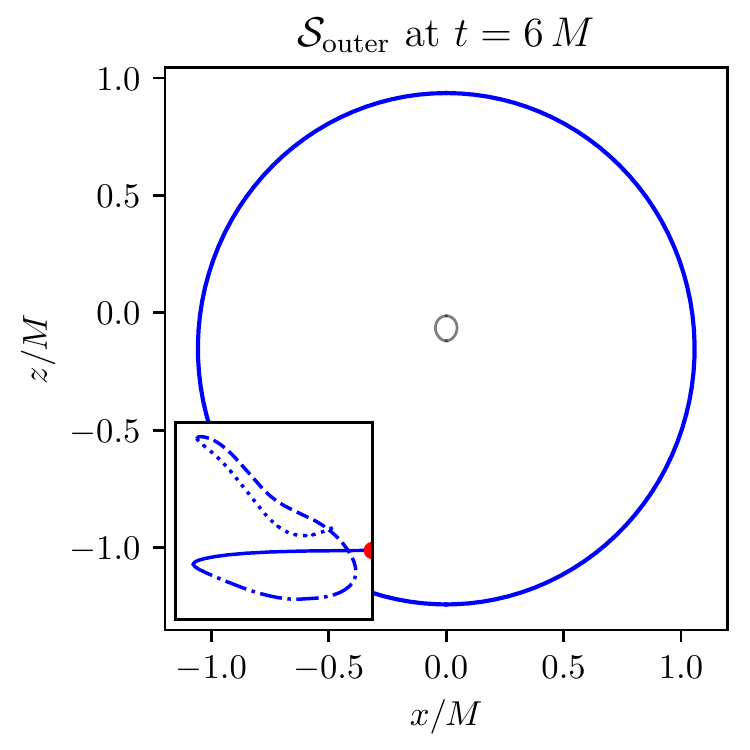}%
    \includegraphics[width=0.333\linewidth]{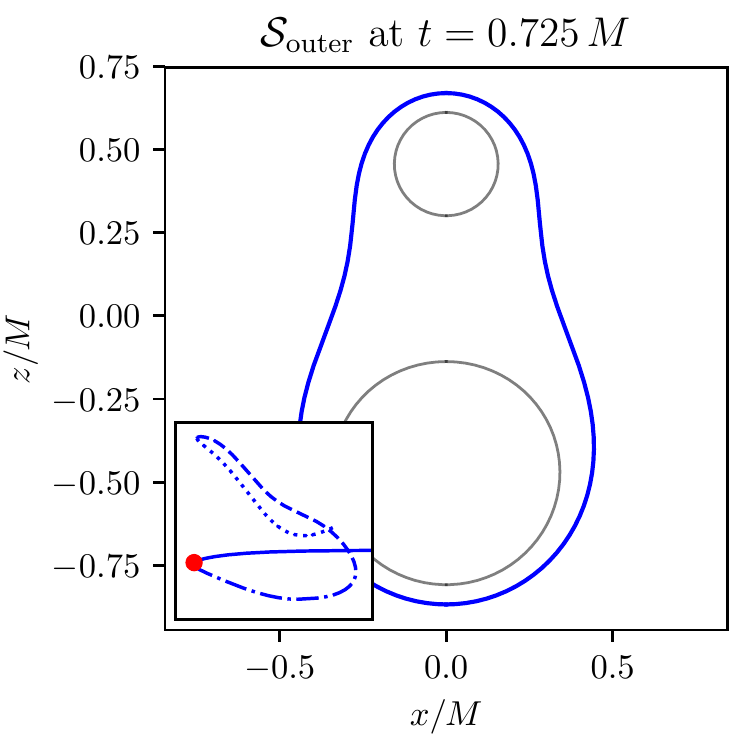}%
    \includegraphics[width=0.333\linewidth]{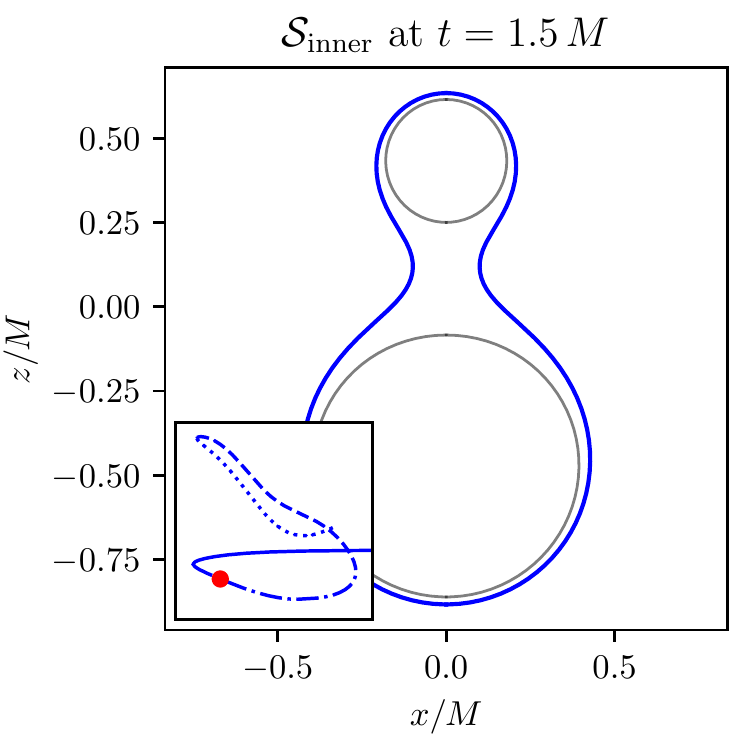}\\
    \includegraphics[width=0.333\linewidth]{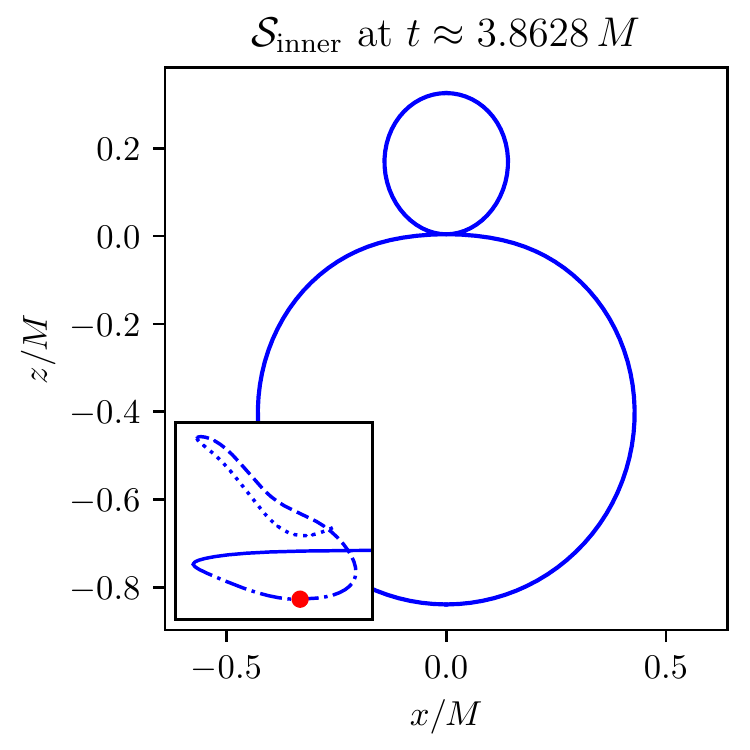}%
    \includegraphics[width=0.333\linewidth]{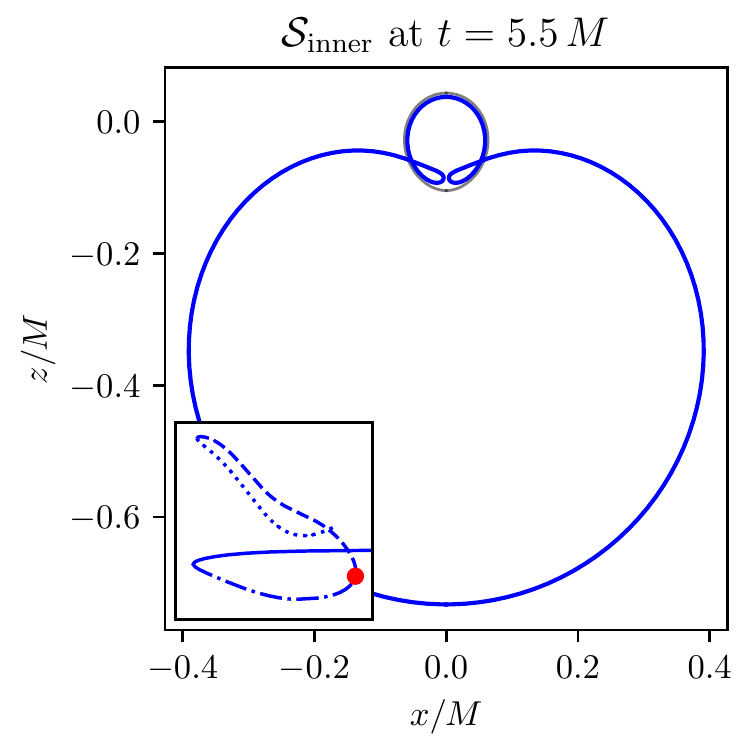}%
    \includegraphics[width=0.333\linewidth]{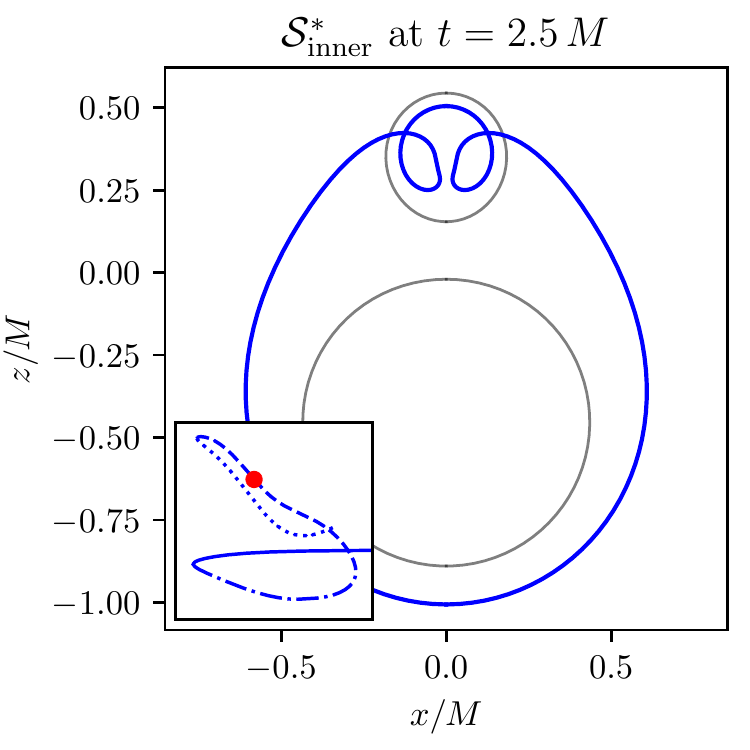}\\
    \includegraphics[width=0.333\linewidth]{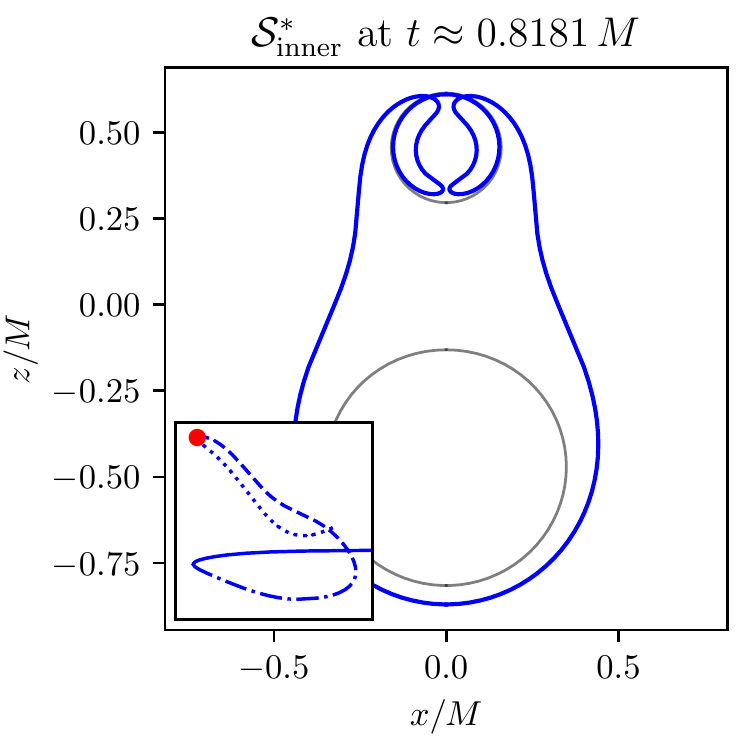}%
    \includegraphics[width=0.333\linewidth]{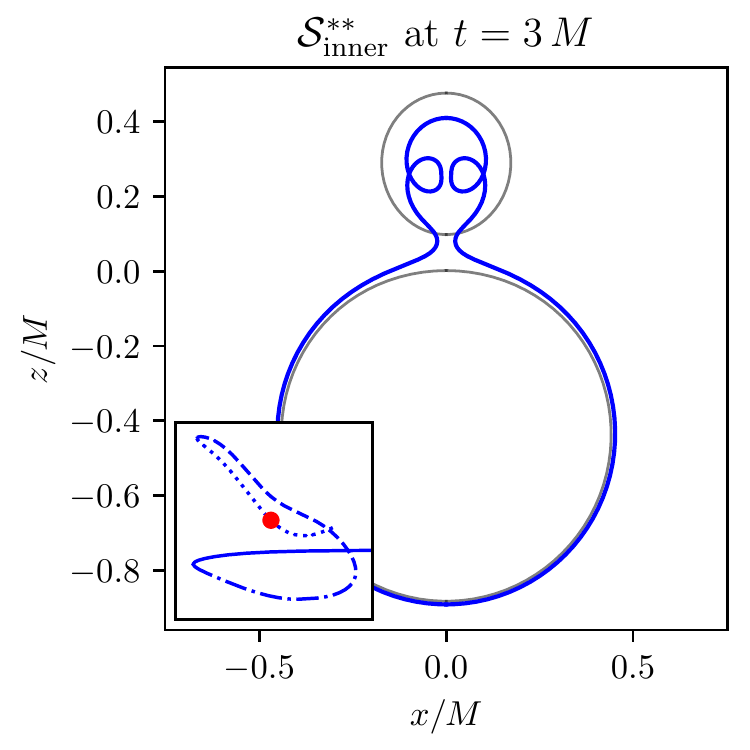}%
    \includegraphics[width=0.333\linewidth]{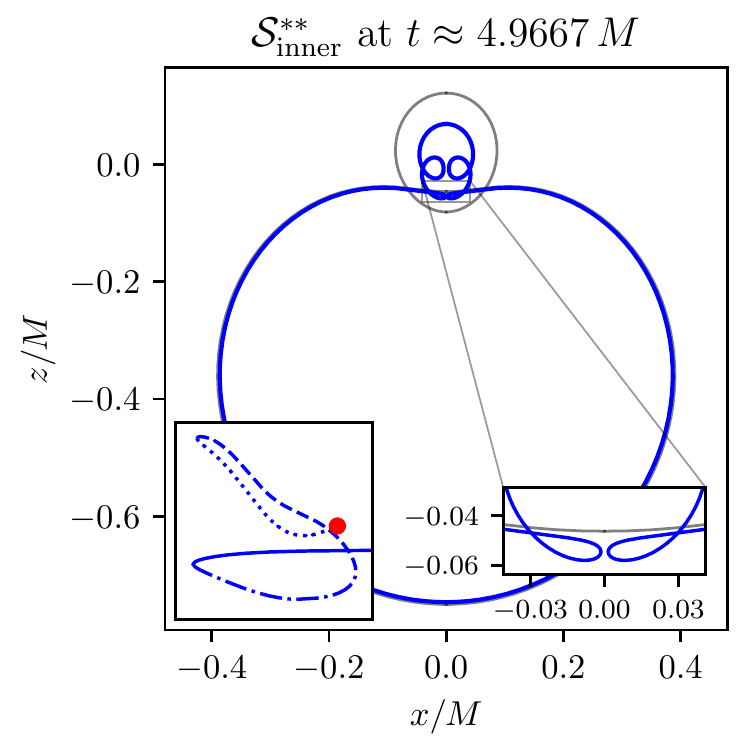}%
    \caption{\label{fig:evolution_outer}%
        Several shapes of the \MOTSs along the connected MOTT following the
        sequence $\Sout \to \Sin \to \Sini \to \Sinii$.
        The panels are to be read row by row from left to right.
        The respective inset in the bottom-left indicates the location of the
        shown MOTS on the connected area curve with a red dot
        (see FIG.~\ref{fig:BL31_areas_overview} for the precise axes and
        labels).
        For reference, the shapes of $\Sone$ and $\Stwo$ are drawn as light
        grey solid lines (except for the top-left and the center panel, where
        $\Stwo$ does not exist).
        We start in the top-left panel with $\Sout$ at $t=6.5M$ and then go
        backwards along the MOTT until we reach $\Sinii$ at the final time
        $t=4.4625M$ when it could be located (bottom-right panel).
        The inset in the bottom-right of this last panel shows the newly
        formed second self-intersection of $\Sinii$.
    }
\end{figure*}

\begin{figure*}
    \includegraphics[width=0.333\linewidth]{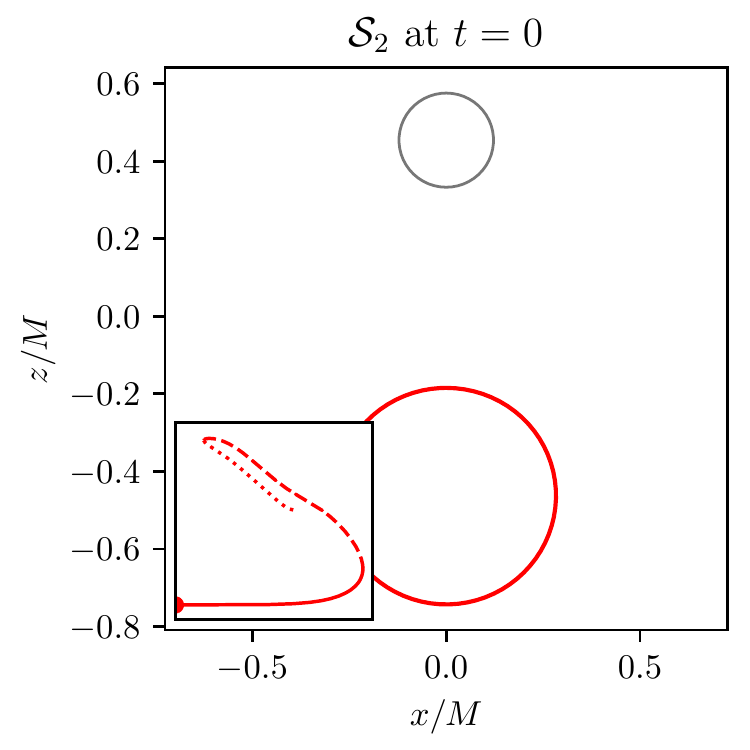}%
    \includegraphics[width=0.333\linewidth]{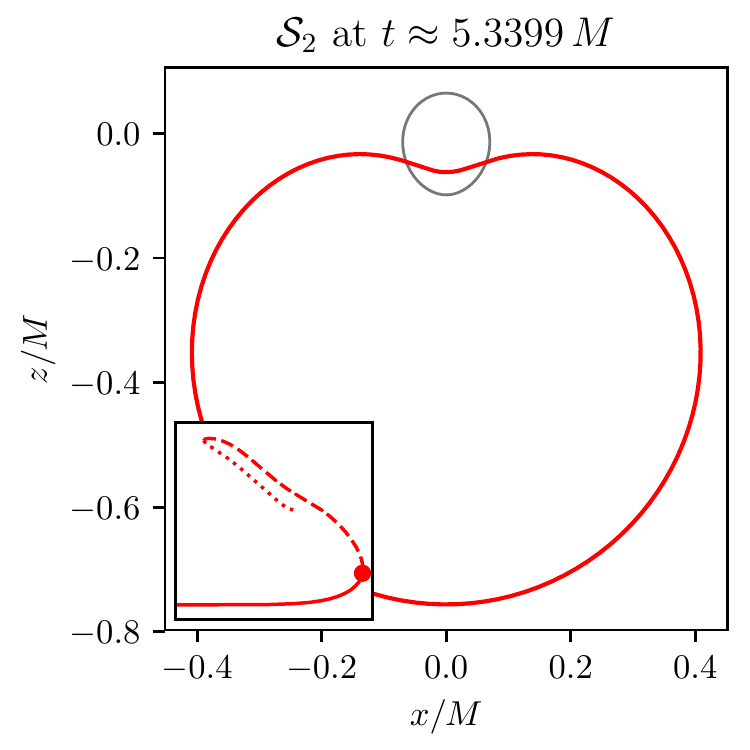}%
    \includegraphics[width=0.333\linewidth]{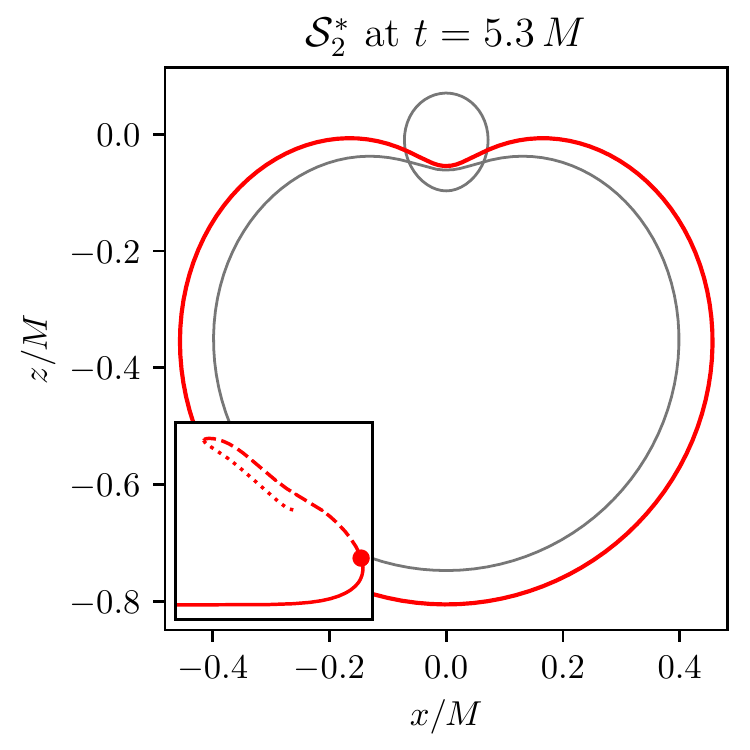}\\
    \includegraphics[width=0.333\linewidth]{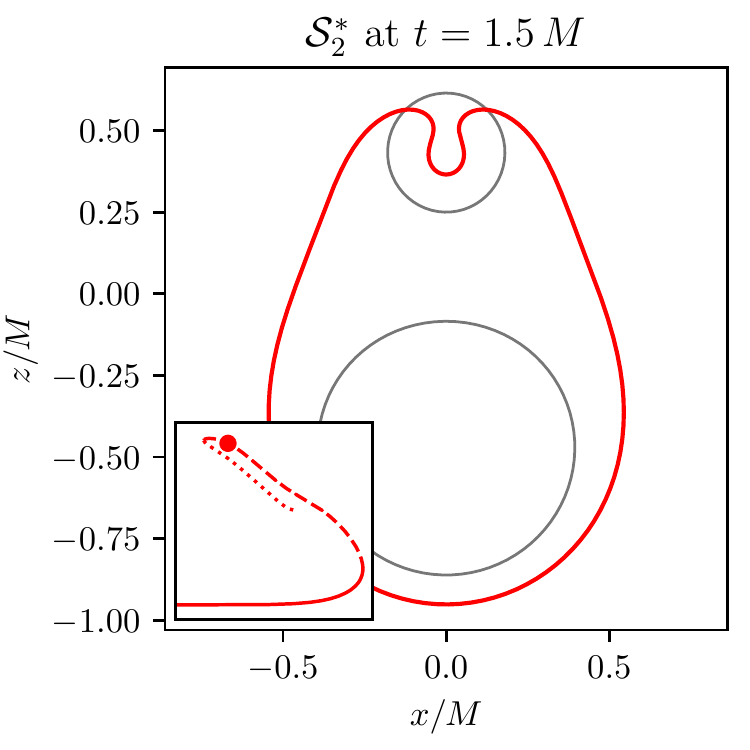}%
    \includegraphics[width=0.333\linewidth]{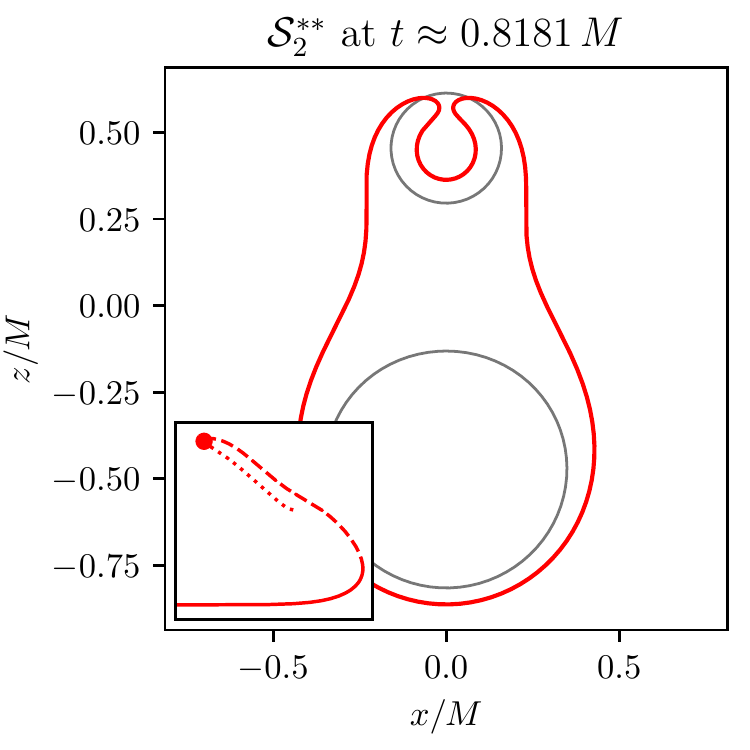}%
    \includegraphics[width=0.333\linewidth]{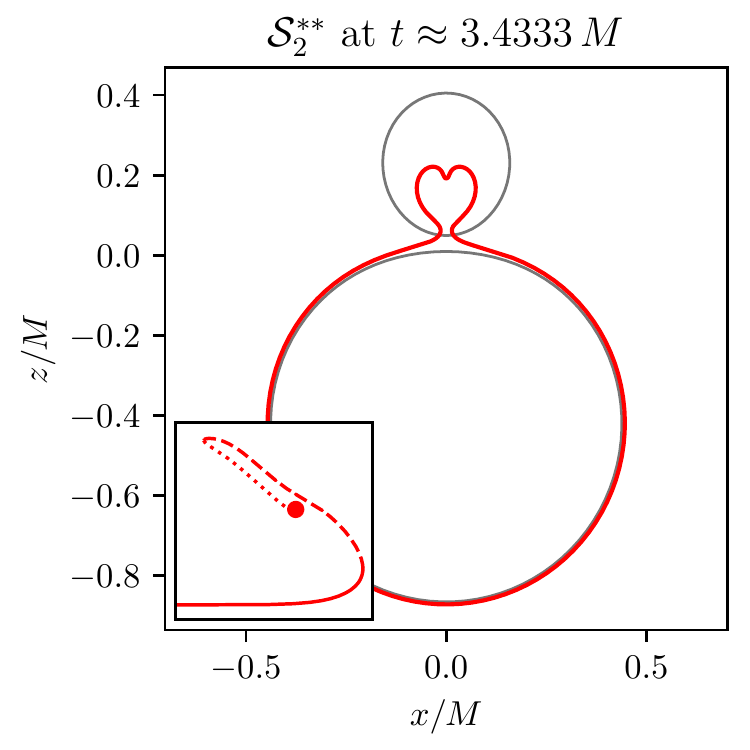}%
    \caption{\label{fig:evolution_S2}%
        Same as FIG.~\ref{fig:evolution_outer} but showing \MOTSs along the
        connected MOTT consisting of $\Stwo \to \Sthree \to \Sthreei$.
    }
\end{figure*}

The most complicated of the four world tubes is the one that
asymptotes to the final Schwarzschild horizon.
Starting with $\Sout$, this MOTT is composed of the sequence
$\Sout \to \Sin \to \Sini \to \Sinii$.
All \MOTSs along this MOTT enclose both punctures.
FIG.~\ref{fig:evolution_outer} shows several examples of these \MOTSs as we follow this
world tube, starting with $\Sout$ (top left panel) and moving backwards.
We find the well-known bifurcation of $(\Sout, \Sin)$ at $\tAH \approx \tAHval$
(top center panel).
Going now forward in time along the inner common MOTT $\Hin$, we find that $\Sin$
coincides momentarily with
$\Sone \cup \Stwo$ at $\ttouch \approx \ttouchval$ (middle left panel)
and afterwards develops self-intersections.
It merges and annihilates smoothly with
$\Sini$, which retains the self-intersection, and travels from the
annihilation backwards in time until shortly after $\tAH$ (bottom left panel).
At this point, it smoothly connects to $\Sinii$ in a bifurcation.
$\Sinii$, initially fully outside and enclosing $\Stwo$, subsequently
becomes increasingly distorted. Just like $\Sin$, it forms a cusp at
the time $\tcuspinii$ when the lower part of it passes from the outside to the
inside of $\Stwo$, after which it has a second self-intersection
(bottom right panel).

The two instances where a cusp is formed along this MOTT are
very similar in nature.
An important ingredient for understanding these is
the maximum principle for \MOTSs (c.f. Section~3.2 in \cite{Moesta:2015sga}). 
This implies  that two smooth \MOTSs $\Surf$ and $\Surf'$, one
enclosing the other, must be identical if they have a common point with
normals pointing in the same direction.
This is precisely the situation of $\Sin$ as it approaches $\ttouch$ and that
of $\Sinii$ as $\tname\to\tcuspinii$.
At the respective times of cusp formation, the two portions of the MOTS
separated by the cusp are individually each a smooth MOTS.
In the case of $\Sin$, this is just $\Sone$ and $\Stwo$, while for $\Sinii$ it
is $\Stwo$ and another MOTS, $\SoneC$, discussed in Section~\ref{sub:insideS12}.

\subsubsection{The world tube of $\Stwo$}
\label{sub:worldtube_S2}

This MOTT consists of the sequence
$\Stwo \to \Sthree \to \Sthreei$, with \MOTSs shown in
FIG.~\ref{fig:evolution_S2}.
Starting with $\Stwo$ at $t=0$ (top left panel), we find the annihilation
of $\Stwo$ with $\Sthree$ at $\tname \approx \tStwoval$.
$\Sthree$ is then followed backwards until the point where it bifurcates with
$\Sthreei$ at $\tname \approx \tSthreeval$ (bottom center panel).
All \MOTSs along this MOTT enclose only the puncture in the interior of $\Stwo$,
i.e. that of the larger of the two original black holes.
The bottom right panel of FIG.~\ref{fig:evolution_S2} shows the last time we
were able to reliably locate $\Sthreei$.
Shortly after this time, it gets too close (in coordinates) to the
puncture inside $\Sone$, resulting in loss of numerical accuracy.
However, just as for $\Sin$ and $\Sinii$, $\Sthreei$ approaches $\Stwo$ from
the outside.
We suspect that $\Sthreei$ will subsequently move to the inside of $\Stwo$ by
forming a cusp at the time of transition. At this time, it must momentarily
coincide with $\Stwo$ in its lower portion and another MOTS
($\SoneA$, shown in Section~\ref{sub:insideS12})
in its upper portion.
If true, this scenario would lead to a self-intersection forming in this
world tube.

The annihilation of $\Stwo$ with $\Sthree$ resolves the previously unknown fate of this individual
apparent horizon, although the fate of the full world tube (including $\Sthree$ and
$\Sthreei$) is not numerically resolved.
We shall defer further discussion of possible scenarios to
Section~\ref{sec:conclusions}.

\subsubsection{The world tube(s) of $\Sone$ and $\Sfour$}
\label{sub:worldtube_S1}

\begin{figure}
    \includegraphics[width=0.9\linewidth]{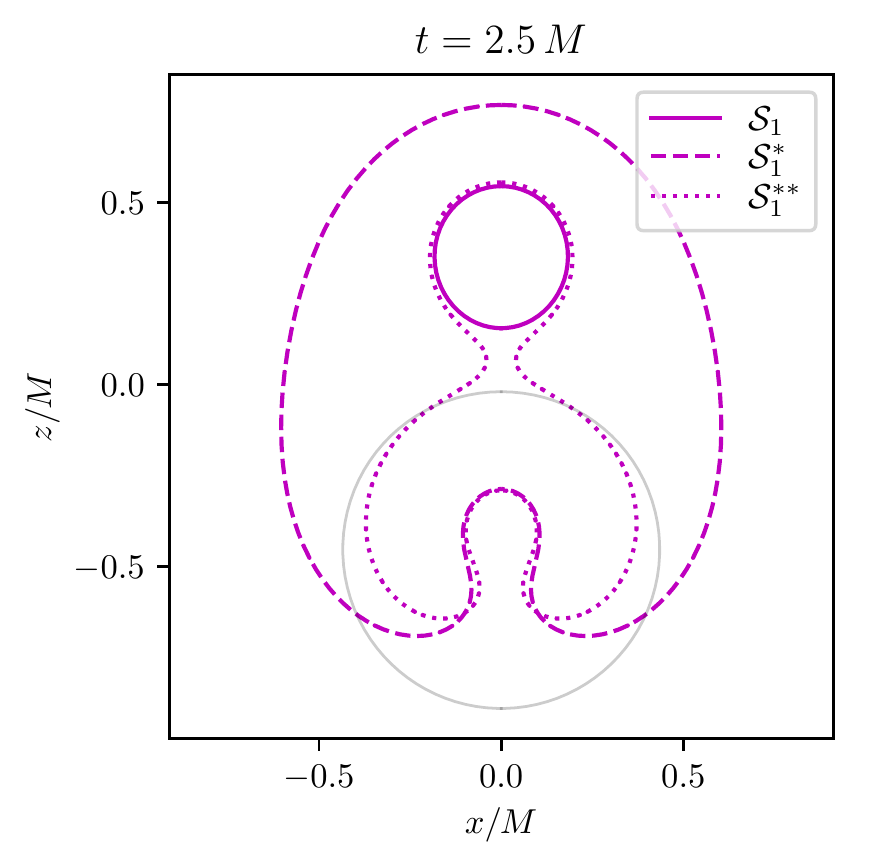}%
    \caption[]{\label{fig:S1_S4_S4i}%
        Examples of the \MOTSs $\Sone$, $\Sfour$, and $\Sfouri$ at simulation
        time $t=2.5M$. The light grey curve shows $\Stwo$ for reference.
    }
\end{figure}

At $\tSfour \approx \tSfourval$ we find the
formation of the pair $(\Sfour, \Sfouri)$.
These enclose the puncture inside $\Sone$ but do not contain the
puncture of $\Stwo$.
FIG.~\ref{fig:S1_S4_S4i} shows this pair along with $\Sone$ at a time $t=2.5M$.

A merger and annihilation of $\Sone$ with $\Sfour$ analogous to that of
$\Stwo$ with $\Sthree$ could not be seen in this simulation.
It is unclear if this lack of an annihilation is
a) purely due to the smaller $\Sone$ being closer in coordinates to one of
the punctures, which increases the effect of the slow-down discussed above and
in \cite{evans:2020lbq}, or if
b) $\Sone$ has a qualitatively different behavior than $\Stwo$ and
possibly does not annihilate at all.
One option to pursue the first possibility is to choose initial data for a
simulation which reduces the aforementioned slow-down effects.
Appendix~\ref{app:S1annihilation} shows results for such a simulation,
which suggests that the lack of annihilation could indeed be due to case a)
even in the present simulation.

Lastly, we would like to point out that, in a very similar way to $\Sthreei$, we find 
the inner branch $\Sfouri$ approaching one of the individual apparent horizons, this
time $\Sone$, from the outside.
In this case, however, we were able to resolve the formation of a cusp as
$\Sfouri$ coincides with $\Sone \cup \StwoA$
($\StwoA$ is discussed in Section~\ref{sub:insideS12}).
As expected, $\Sfouri$ subsequently self-intersects.
This supports again the above expectation that this also happens for $\Sthreei$.

\subsubsection{The world tube of $\Sfive$}
\label{sub:worldtube_S0}

\begin{figure}
    \includegraphics[width=0.9\linewidth]{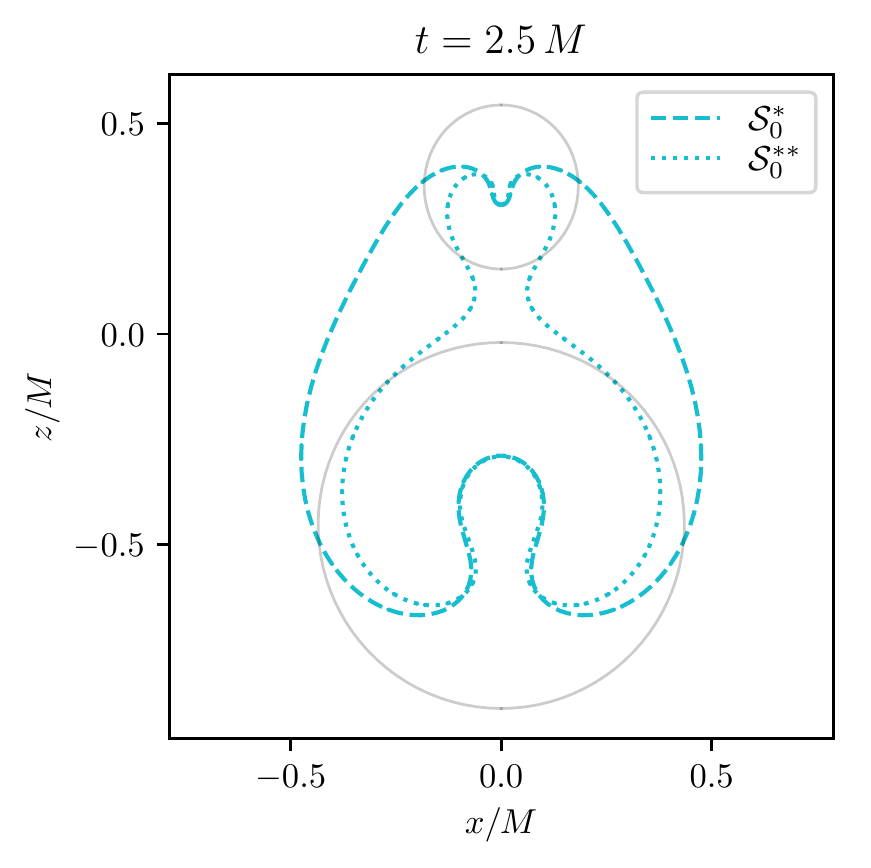}%
    \caption[]{\label{fig:S5_S5i}%
        Examples of the \MOTSs $\Sfive$ and $\Sfivei$ at simulation
        time $t=2.5M$. The light grey curves show $\Sone$ and $\Stwo$ for
        reference.
    }
\end{figure}

The last world tube is traced out by the pair $(\Sfive, \Sfivei)$,
which bifurcates at $\tSfive \approx \tSfiveval$.
These do not connect smoothly to any of the above \MOTSs.
FIG.~\ref{fig:S5_S5i} shows the \MOTSs at a time $t=2.5M$.
None of the \MOTSs along this world tube contain any puncture
in their interior.

Based on the cases of cusp formation and self-intersections
mentioned above, we speculate here on the future of $\Sfivei$ beyond the point
where we are able to locate it reliably:
We already mentioned the cusp formations of $\Sfouri$ and $\Sthreei$ in
the previous two subsections. In these two cases, either $\Sone$ or $\Stwo$
coincides with $\Sfouri$ or $\Sthreei$ on one of their portions, respectively.
We also noted that in both cases the two remainders,
$\StwoA$ and $\SoneA$, respectively,
are themselves \MOTSs discussed in Section~\ref{sub:insideS12}.
These latter two \MOTSs do not contain any puncture.
They may, however, at some point touch and start to intersect, just like
$\Sone$ and $\Stwo$ do at $\ttouch$.
At the time when they touch, we propose that $\Sfivei$ coincides with their
union, has a cusp at this time, and forms a self-intersection immediately
afterwards.
We are unfortunately not able to resolve this idea at this point since we lose
numerical accuracy near the punctures before this can be observed.

\subsubsection{Comparison with the extreme mass ratio merger case}
It is illuminating to compare these sequences of \MOTSs to FIG.~16 in \cite{Booth:2020qhb}. 
That paper studied marginally outer trapped open surfaces (MOTOSs) in Schwarzschild spacetimes
with the argument that they could be used to model dynamical apparent horizon and other MOTT evolutions
during an extreme mass ratio merger. 

That paper did not include true dynamics, but it was argued that it should be possible to assemble the various MOTOSs found 
in Schwarzschild into a sequence describing the full dynamics of the merger. The authors proposed one such evolution, but what they 
didn't consider\footnote{We are confident about this  as two of them are also authors on the current paper!} is that many of the steps could be happening 
simultaneously. Instead of creations and annihilations, they expected a single continuous sequence of MOTOSs evolving in time. If one 
reinterprets the sequence in their FIG.~16 as a marginally outer trapped open tube weaving backwards and forwards in time, then the proposed
picture becomes much closer to that of FIG.~\ref{fig:evolution_outer} in the current paper, though still with some mistakes. 

It is intriguing that that simple model of an extreme mass ratio merger can reproduce cusp formation which subsequently
evolves into self-intersections. However since  \cite{Booth:2020qhb} works with an exact solution, we can push further and see that this process 
may repeat indefinitely with surfaces involving more and more self-intersections. This then is extra evidence that the double self-intersection of 
FIG.~\ref{fig:evolution_outer} at $t\approx 4.9667M$ is likely just the beginning of a sequence that may continue indefinitely with each $n$-times
self-intersecting MOTOS ultimately being annihilating with an $(n+1)$-times self-intersecting MOTS which in turn was pair-created with a (soon-to-be) $(n+2)$-times self-intersecting MOTOS.

\subsection{\MOTSs inside $\Sone$ and $\Stwo$}
\label{sub:insideS12}

\begin{figure*}
    \includegraphics[width=1.0\linewidth]{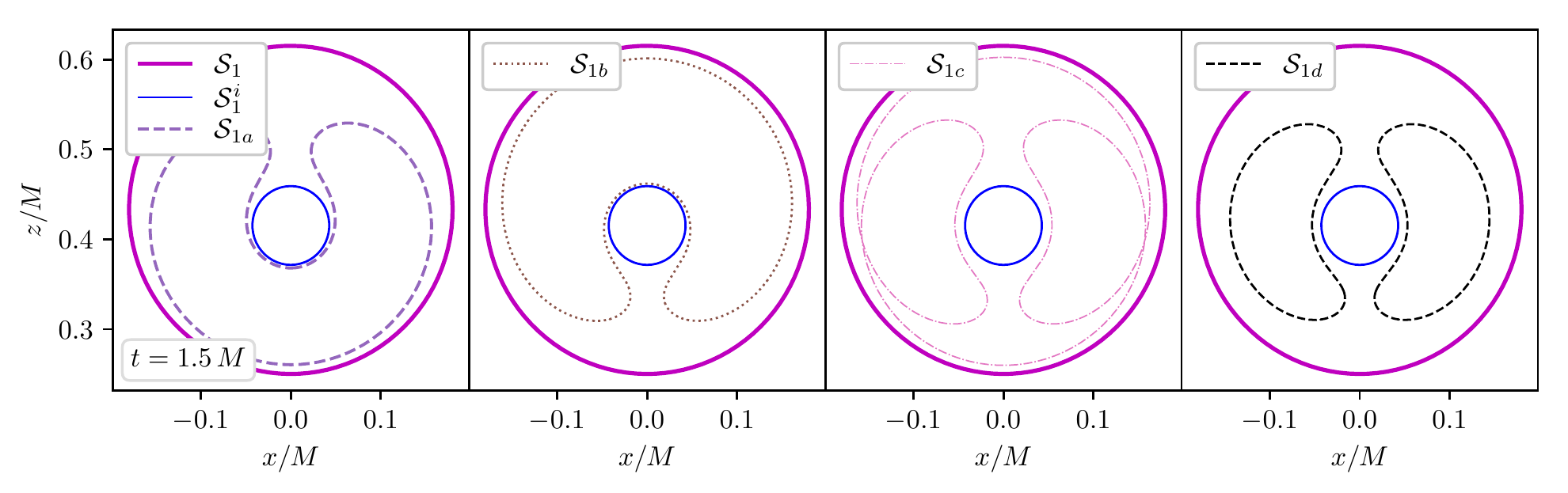}%
    \caption[]{\label{fig:MOTSs_inside_S1}%
        The \MOTSs $\SoneABCD$ in the interior of $\Sone$ at a simulation
        time $\tname=1.5M$.
        $\SoneA$ (first panel) and $\SoneB$ (second panel) do not enclose the
        puncture inside $\Sone$, while $\SoneC$ (third panel) is
        self-intersecting and does enclose this puncture.
        The last panel shows a MOTS $\SoneD$ of toroidal topology.
        We also show a marginally {\em inner} trapped surface $\iSone$ as
        thin solid line, which is discussed in Section~\ref{sub:MITS}.
    }
\end{figure*}

\begin{figure*}
    \includegraphics[width=0.48\linewidth]{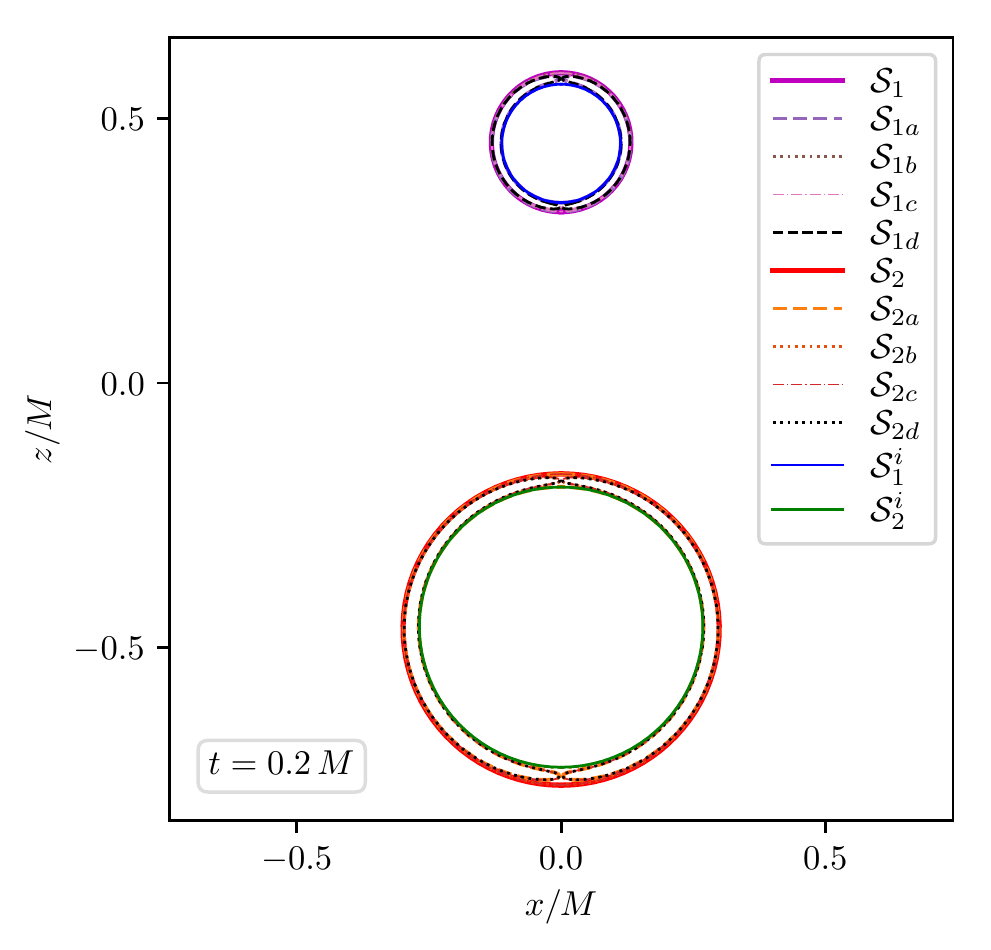}\hfill%
    \includegraphics[width=0.48\linewidth]{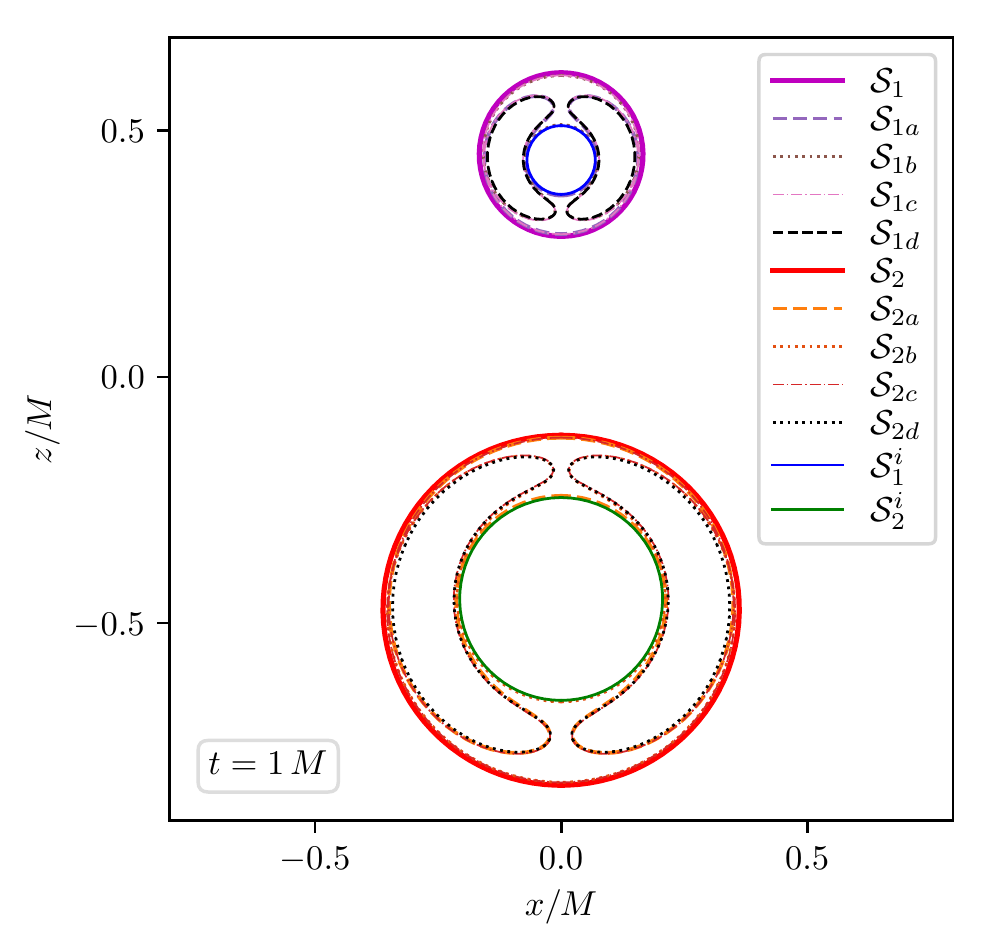}%
    \caption[]{\label{fig:MOTSs_inside_S12}%
        \MOTSs in the interior of the individual \MOTSs $\Sone$ and $\Stwo$
        at the two different times $t=0.2 M$ (left panel)
        and $t=1M$ (right panel).
        Any MOTS outside or partially outside the individual ones is not shown here.
        The \MOTSs $\SoneABCD$ are shown individually in
        FIG.~\ref{fig:MOTSs_inside_S1}.
        We furthermore show the \MITSs $\iSone$, $\iStwo$, which are discussed in
        Section~\ref{sub:MITS}.
    }
\end{figure*}

None of the \MOTSs described thus far are located fully in the interior of
either $\Sone$ or $\Stwo$.
As discussed in the first paper, we were in fact not able to
locate such an interior MOTS in the Brill-Lindquist initial data.
From the discussion in the previous Section~\ref{sub:worldtubes}, however,
one would expect that at any point where a cusp forms in a MOTS and separates
it into two smooth parts, these two parts
are individually separate \MOTSs with their own
time development to the future and past.
This holds for example for the upper portion of $\Sinii$ at the time it forms
a cusp shortly before the last panel of FIG.~\ref{fig:evolution_outer}.
This portion is fully contained inside $\Sone$, i.e. at some point a MOTS must
have formed inside $\Sone$ evolving into this self-intersecting shape.
A similar argument holds for the lower portion of $\Sfouri$.

\begin{figure*}
    \includegraphics[width=0.48\linewidth]{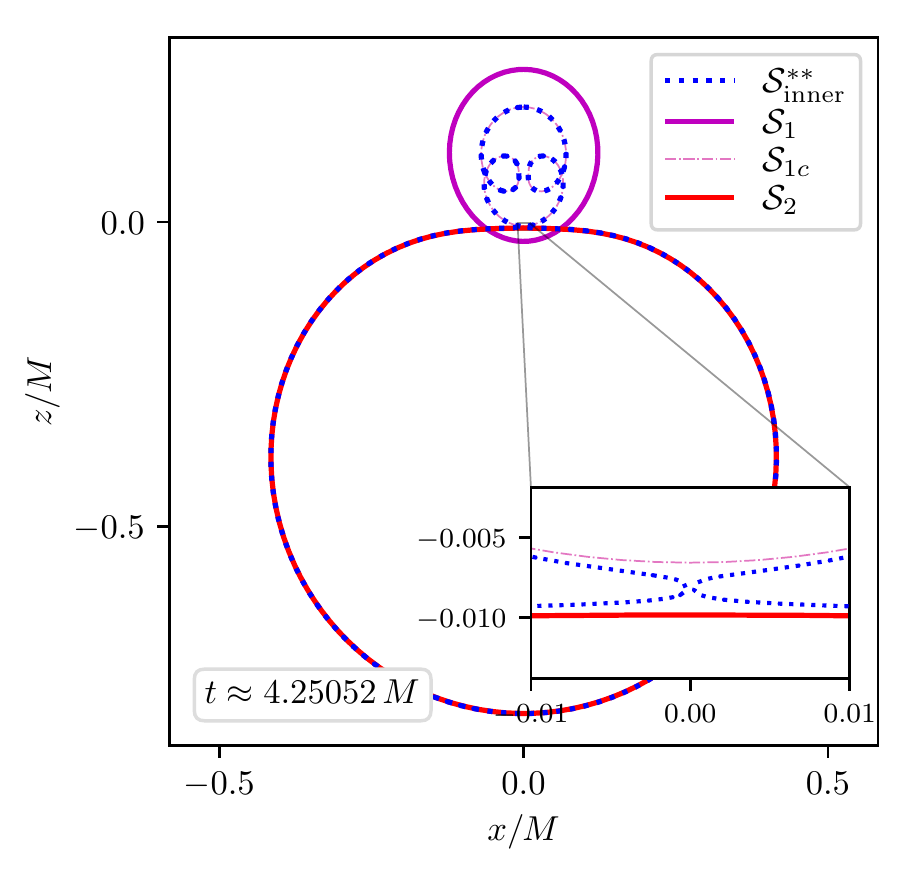}\hfill%
    \includegraphics[width=0.48\linewidth]{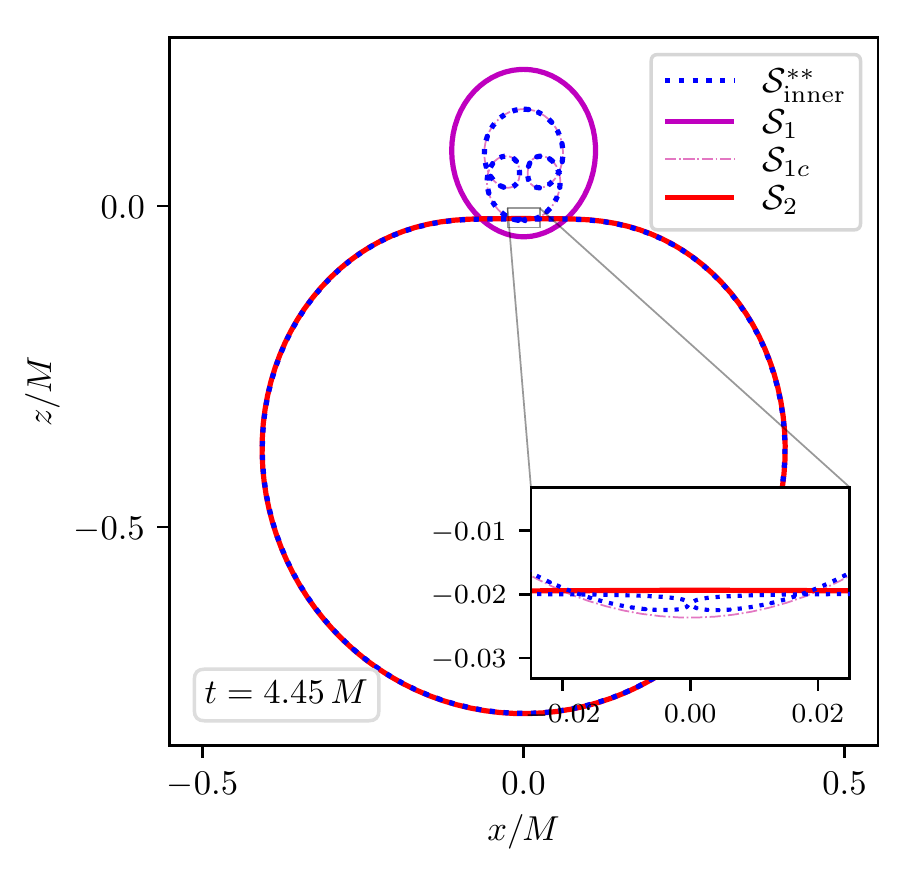}%
    \caption[]{\label{fig:MOTSs_tcusp}%
        Cusp formation with subsequent self-intersection in the evolution of
        $\Sinii$. The left panel shows a time before cusp formation and the
        right panel a time after. Note that the cusp forms at the time when
        $\SoneC$ and $\Stwo$ touch. The interior self-intersecting \MOTSs are very 
        similar to those see in pure Schwarzschild in  \cite{Booth:2020qhb}.
    }
\end{figure*}

A search for \MOTSs in the interior regions of $\Sone$ and $\Stwo$ at a time
$t>0$ has indeed been successful.
FIG.~\ref{fig:MOTSs_inside_S1} shows the four \MOTSs we could locate inside
$\Sone$ at a simulation time $t=1.5M$.
From FIG.~\ref{fig:MOTSs_inside_S12} it can be seen that these \MOTSs pinch
off and merge with $\Sone$ as $t \to 0$, i.e. they do not exist separately in
the initial data and thus can only be found at later times.
The same qualitative behavior is found inside $\Stwo$.
Note that the self-intersecting MOTS $\SoneC$ later merges with $\Sinii$ at
the time the latter forms the cusp at $\tname\approx\tcuspiniival$.
FIG.~\ref{fig:MOTSs_tcusp} shows this formation of a cusp and subsequent
self-intersection.
Similarly, $\StwoA$ merges with $\Sfouri$ as it forms its cusp at about
$\tname\approx\tcuspfourival$.
Another observation is that these interior \MOTSs cannot ``escape'' their
enclosing $\Sone$ or $\Stwo$, respectively, while the latter exist.
This is again easily explained by the maximum principle for \MOTSs
given in Section~3.2 of Ref.~\cite{Moesta:2015sga}, since any common point
with a common normal direction would imply the \MOTSs necessarily coincide.

Interestingly, $\SoneD$ and $\StwoD$ are \MOTSs with toroidal topology,
i.e. their Euler characteristic is $\chi=0$ in contrast to all other \MOTSs
which have $\chi=2$.
We numerically verified that their shear $\sigma^+_{AB}$ does not vanish,
which immediately implies that any first order spacelike outward deformation
does not lead to an untrapped surface
(see the remark at the end of Section~III~A of Ref.~\cite{Ashtekar:2003hk}).
This is compatible with a negative principal eigenvalue of the stability
operator, which we find as well.
$\SoneD$ is shown in the last panel of FIG.~\ref{fig:MOTSs_inside_S1} and
closely follows the loop of the self-intersecting $\SoneC$.
Just like $\SoneABC$ and $\StwoABC$, the toroidal \MOTSs merge with $\Sone$
and $\Stwo$, respectively, as $t \to 0$.

\begin{figure*}
    \includegraphics[width=0.48\linewidth]{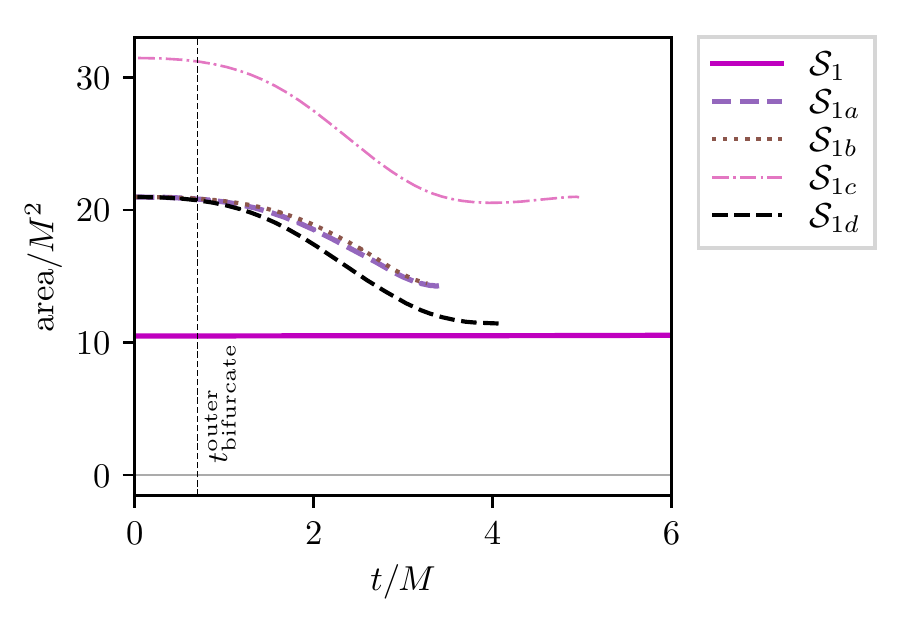}\hfill
    \includegraphics[width=0.48\linewidth]{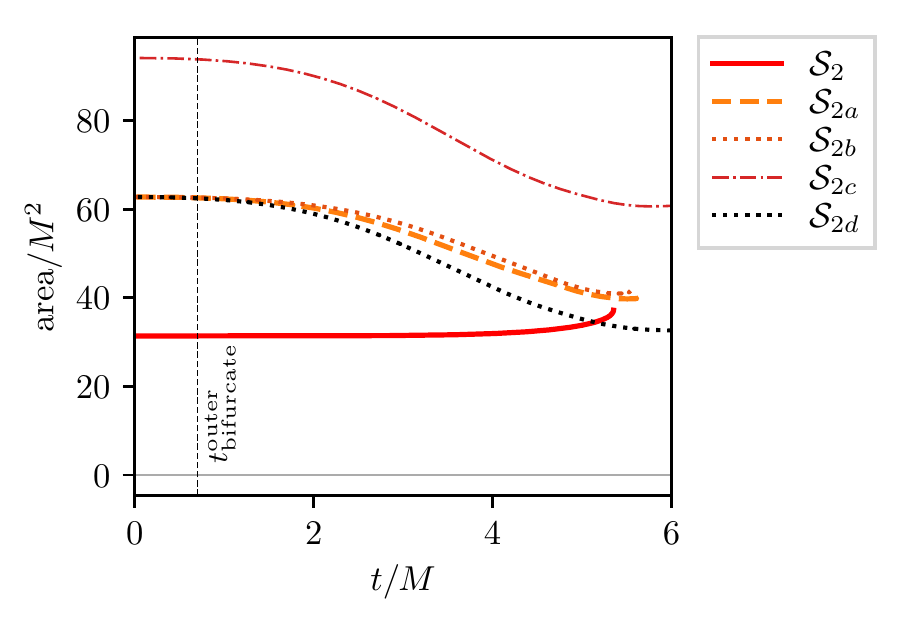}%
    \caption{\label{fig:BL31_areas_inside_S12}%
        Evolution of the area of the \MOTSs $\SoneABC$ inside $\Sone$
        (top panel) and
        of $\StwoABC$ inside $\Stwo$ (bottom panel).
        The areas of $\Sone$ and $\Stwo$ are shown for reference.
        Note that the smooth annihilation of $\Stwo$ is not visible since we
        do not include the area of the connecting MOTS $\Sthree$.
    }
\end{figure*}

From their behavior for $t \to 0$, one can immediately deduce that the area of
$\SoneAB$ and $\SoneD$ becomes twice the area of $\Sone$
while $\SoneC$ has three times this area in the limit $t \to 0$.
Analogous arguments hold for $\StwoABCD$.
FIG.~\ref{fig:BL31_areas_inside_S12} confirms this expectation and shows the
evolution of the various areas as a function of time.

All these interior \MOTSs get very close in coordinates to one of the
punctures and we hence lose most of them due to numerical problems before the
end of our simulation.
This happens earlier for $\SoneABCD$ than for $\StwoABCD$ as the former are
smaller in coordinates than the latter.

\subsection{\MOTSs seen from the other asymptotic ends}
\label{sub:MITS}

Figures~\ref{fig:MOTSs_inside_S1} and \ref{fig:MOTSs_inside_S12} show curves
$\iSone$ and $\iStwo$ which we did not yet comment on.
These curves belong to marginally {\em inner} trapped surfaces (\MITSs), where
$\Theta_{-} = 0$ with no condition on $\Theta_{+}$.
Equivalently, these surfaces are \MOTSs with the notion of {\em inward} and
{\em outward} reversed.
These \MITSs are interesting for two reasons.

\begin{figure}
    \includegraphics[width=0.9\linewidth]{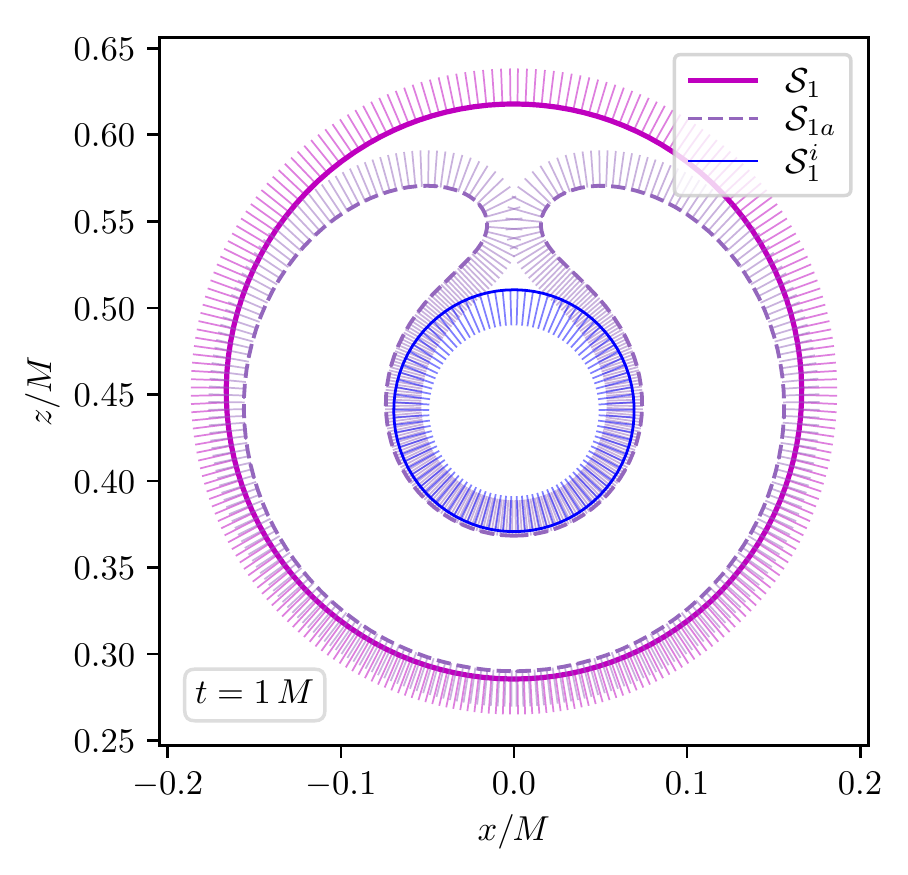}%
    \caption{\label{fig:S1a_normals}%
        Spatial outward normal directions on $\Sone$ and one of the interior
        \MOTSs $\SoneA$ as well as the inward normal direction on the MITS
        $\iSone$.
    }
\end{figure}

First, the \MITSs provide a means for generalizing a feature visible in the
time-symmetric case (see paper I), where the \MOTSs with
sharp turns seem to transition between portions staying
close to one of the other \MOTSs with fewer or no turns.
However, at these turns the notion of inside and outside may switch.
This is irrelevant in time-symmetry as in that case
$\Theta_{+} = 0 \Longleftrightarrow \Theta_{-} = 0$,
but it becomes important during the simulation where time-symmetry is lost.
As can be seen in FIG.~\ref{fig:MOTSs_inside_S1}, the \MOTSs $\SoneABCD$
run close to $\iSone$ on an extended portion,
consistent with the notions of inside and outside there.
FIG.~\ref{fig:S1a_normals} visualizes this idea by showing the spatial part of
the respective null normals with vanishing expansion, i.e. of $\ell^+$ for the
\MOTSs and $\ell^-$ for the MITS.
We find that all \MOTSs in the interior of $\Sone$ fully lie in
the annular region between $\Sone$ and the MITS $\iSone$.
The \MOTSs in the interior of $\Stwo$ behave analogously.

\begin{figure}
    \includegraphics[width=0.9\linewidth]{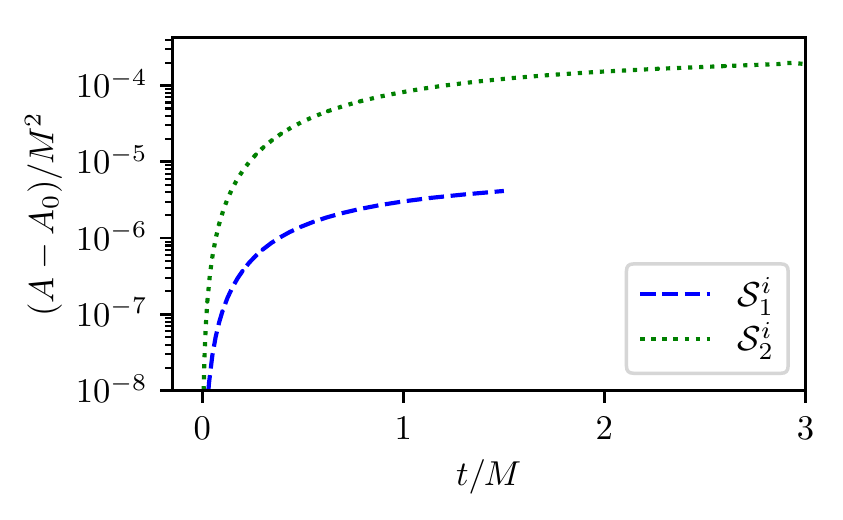}%
    \caption{\label{fig:MITSs_area}%
        Evolution of the area of the \MITSs $\iSone$ and $\iStwo$.
        We plot the difference $A-A_0$ on a logarithmic scale to emphasize the
        area change. Here $A_0$ is the area of the MITS at $t=0$ where it
        coincides with the area of $\Sone$ and $\Stwo$, respectively.
    }
\end{figure}

The second reason to consider \MITSs is that the two punctures in the
Brill-Lindquist initial data belong to two different asymptotically flat ends
of the slice $\Sigma$.
In fact, these data contain three ends with $x \to \infty$
representing the end we commonly choose to be in the ``outward'' direction.
The ends $x \to x_1$ and $x \to x_2$, where $x_{1,2}$ are the coordinates of
the two punctures, contain equally valid observers far away from any
black hole.
An observer near $x_1$ or $x_2$ will see a quite different picture of what
``we'' (i.e. observers for whom $x\to\infty$ is considered outside)
see as a binary black hole merger.
In particular, at $t = 0$ where $\Sonetwo$ are not only \MOTSs but also
\MITSs, an observer near, say, $x_1$ interprets $\Surf_1$ as {\em common}
apparent horizon enclosing both punctures.
FIG.~\ref{fig:MITSs_area} shows that the area of these \MITSs is monotonically
increasing.
Furthermore, considered as \MOTSs seen from the respective asymptotic region,
$\iSone$ and $\iStwo$ are strictly stable
and, in fact, the {\em outermost} common \MOTSs representing a
perturbed Schwarzschild black hole.
While many more \MITSs are likely present in our simulations, we shall leave a
more exhaustive study of these surfaces for future work.

\subsection{Remarks on initial distance and mass ratio}
\label{sub:mass_ratio}

Most of the discussed numerical results are obtained from simulations starting
with a single physical configuration, namely Brill-Lindquist initial data
consisting of a conformally flat three-metric with conformal
factor~\eqref{eq:BLmetric} where the bare masses are
$m_1 = 1/3$ and $m_2 = 2/3$ and the distance parameter is chosen as $d=0.9$.
This choice balances several effects resulting from the interplay of the
employed slicing and the numerical setup:
A larger value for $d$ makes the individual apparent horizons $\Sonetwo$ slow
down in evolution before they start to intersect \cite{evans:2020lbq} and thus
prevents us from observing the annihilations of $(\Stwo, \Sthree)$
and $(\Sin, \Sini)$.
A smaller value leads to common \MOTSs $\Sout$ and $\Sin$ already being present
in the initial data. As shown in paper I, many more common \MOTSs exist in
these kinds of setups and hence at least some of the various bifurcations
do not exist in the simulation.

Similar considerations led to the chosen mass ratio of $q = 2$.
In particular, it is true that more unequal masses keep the larger MOTS
$\Stwo$ further away (in coordinates) from the puncture in its interior.
As a consequence, it suffers less from the slow-down described in
\cite{evans:2020lbq} and we can observe that it annihilates with $\Sthree$
``earlier'' with respect to simulation time.
However, this comes at the cost of $\Sthree$ now getting too close (again, in
coordinates) to the puncture inside the {\em smaller} MOTS $\Sone$ and thus we
were not able to resolve the full world tube of $\Sthree$ and show that it
turns around in time at {\em both} of its ends.
This also means that the prospect of resolving the fate of $\Sone$
crucially depends on keeping it as far away from its puncture as possible.
Without modifying the used gauge conditions, which we do not try to do in
the present work, this can only be achieved with more equal masses.
Our attempts to do so are summarized in Appendix~\ref{app:S1annihilation}.
However, we find that this requires a very small initial distance $d$.

Thus, while we are indeed able to resolve some of the previously presented
features for many choices of initial data, the particular choice we made
combines most of these in one simulation.
Furthermore, the reasons we cannot resolve the full set of these features for
other mass ratios or initial distances are well understood and due to
numerical issues.
We therefore have no clear indication that the observed behavior should indeed
be specific to our choice of these parameters.
The exception is a possible qualitative change in case of a large mass ratio.
For large $q$, we find that the time $\tStwo$ where $\Stwo$ annihilates with
$\Sthree$ occurs earlier and much closer to the time $\ttouch$ when $\Sone$
and $\Stwo$ start to intersect.
Recall that $\ttouch$ is precisely the time when the union $\Sone \cup \Stwo$
coincides with $\Sin$
and that this provides the connection between $\Sonetwo$ and $\Sout$
\cite{PhysRevLett.123.171102,PhysRevD.100.084044}.
We have verified for $d=0.9$ that $\ttouch < \tStwo$ up to $q=14$.
However, if one finds for even larger $q$ that $\Stwo$ annihilates with
$\Sthree$ before it intersects with $\Sone$, one may still be able to find a
sequence of \MOTSs connecting $\Sonetwo$ with $\Sout$.
In this case, the connection may occur after $\Stwo$ has turned around in
time, i.e. one may see a merger $\Sone \cup \Sthree$ with $\Sini$.

\section{Stability}
\label{sec:stabilityResults}

The various bifurcations and annihilations described thus far can also be
understood with the help of the MOTS stability operator.
As we shall see, this not only gives strong numerical support for our claims of
smooth bifurcations and annihilations,
it also provides a useful characterization of the respective two
branches in terms of the eigenvalues of this operator.

\begin{figure}
    \includegraphics[width=0.9\linewidth]{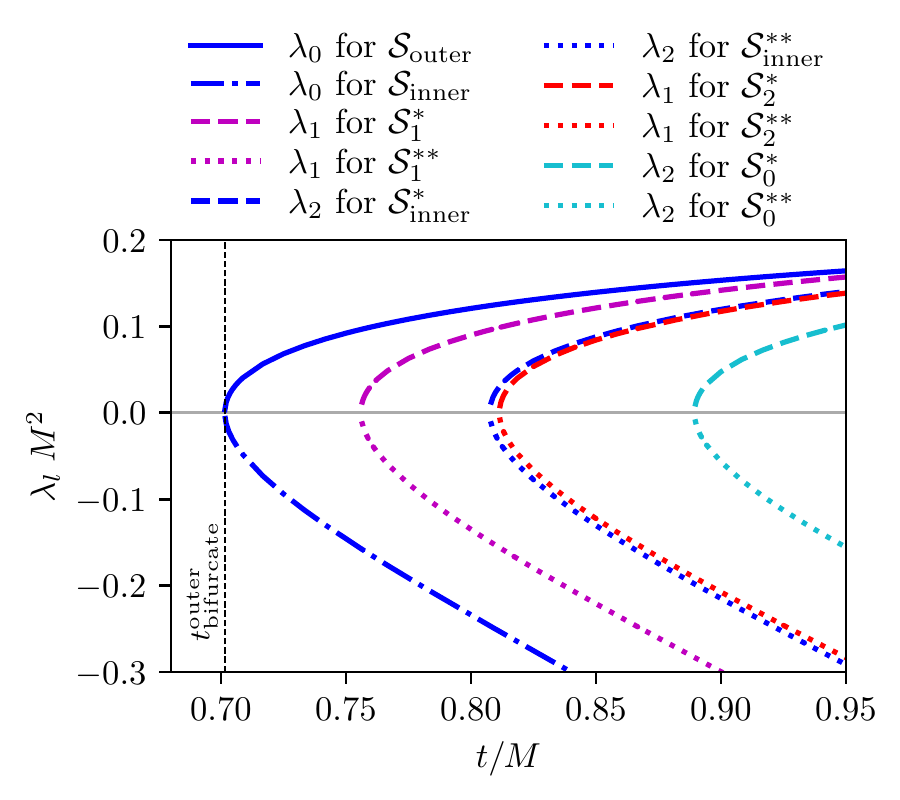}%
    \caption{\label{fig:stability_bifurcations}%
        Eigenvalues of $\Lslice$ for the ten \MOTSs participating in the
        five bifurcations.
        For each MOTS, we show the respective eigenvalue which tends to
        zero as $t \to \tbifurcate$.
    }
\end{figure}

\begin{figure}
    \includegraphics[trim=0 0 0 35,clip,width=0.9\linewidth]{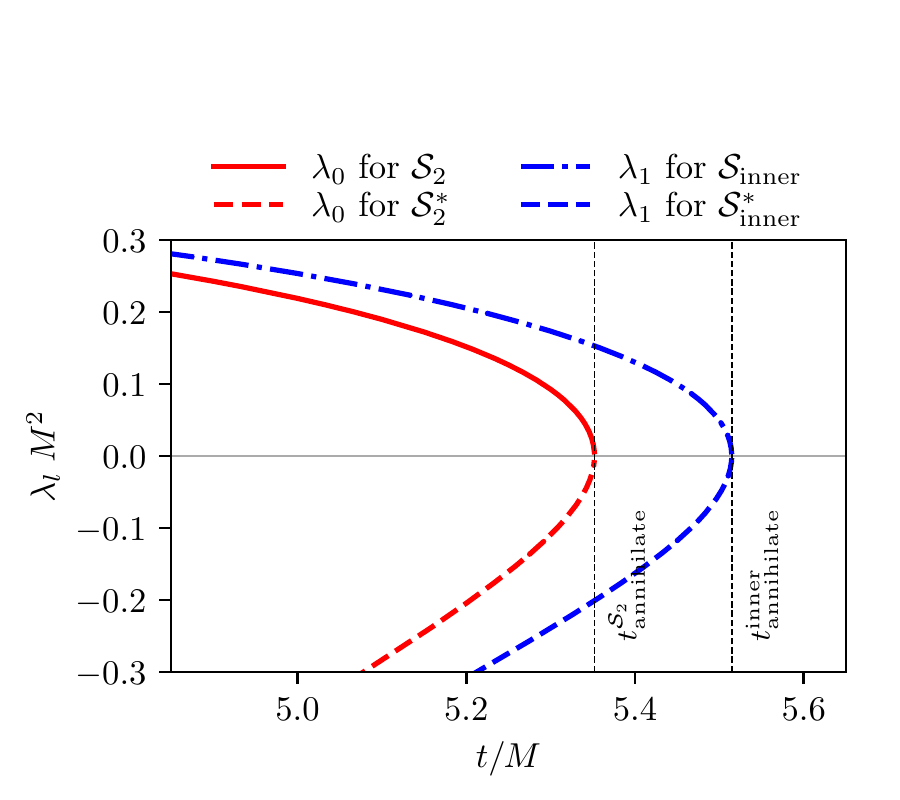}%
    \caption{\label{fig:stability_annihilation}%
        Eigenvalues of $\Lslice$ for the two pairs of \MOTSs
        $(\Stwo, \Sthree)$ and $(\Sin, \Sini)$
        close to the time when they annihilate.
    }
\end{figure}

Let $\Lslice$ be the stability operator \eqref{eq:stab}.
As shown in Proposition~5.1 of \cite{Andersson:2008up}, the vanishing of the
principal eigenvalue of $\Lslice$ is closely related to bifurcations and
annihilations of a MOTS.
The intuitive picture is that of a MOTT $\HH$ which is tangent to one of
the slices $\Sigma_{\tname^*}$, where $\tname^*$ is the time of bifurcation or
annihilation.
At this time, the cross section $\Surf_{\tname^*}$ of $\HH$ is a MOTS with
vanishing principal eigenvalue $\lambda_{0} = 0$.
Essentially, the proposition proves, under suitable genericity conditions
satisfied in all our cases, that the existence of such a MOTS $\Surf_{\tname^*}$
with $\lambda_{0} = 0$ implies existence of a {\em unique} MOTT $\HH$
tangent to $\Sigma_{\tname^*}$ and containing $\Surf_{\tname^*}$.
Hence, if we do find two \MOTTs connecting smoothly at $\tname^*$ with
$\lambda_{0} \to 0$ as $\tname\to\tname^*$, then we have a clear numerical
indication for such a bifurcation or annihilation.
This is precisely the case for the bifurcation of the pair $(\Sout, \Sin)$
and the annihilation of the pair $(\Stwo, \Sthree)$ shown in
Figures~\ref{fig:stability_bifurcations} and \ref{fig:stability_annihilation}.
However, for the other pairs of \MOTSs, we in fact find that instead of the
principal eigenvalue, it is one of the {\em higher} eigenvalues which tends to
zero at bifurcation or annihilation time.

\begin{table}
    \caption{%
        Number of negative eigenvalues of $\Lslice$ for the \MOTSs along the
        different \MOTTs.
        The arrows indicate when we have found a smooth connection between the
        respective world tubes, while the arrow in parentheses indicates a suspected
        smooth transition which we could not resolve numerically.
        Shown is $\NnegO$, i.e. the number of eigenvalues $\lambda_{l,m=0}<0$
        as well as $\Nneg$ for the number of all negative eigenvalues.
        This latter value changes for some of the \MOTTs, in which case we list
        all the occurring cases (not in order of appearance).
    }
    \label{tab:Nneg}
    \begin{ruledtabular}
        \begin{tabular}{lllll}
            MOTT &
                \multicolumn{1}{l@{\ \ \makebox[0pt]{$\!($}$\to$\makebox[0pt]{$\,)$}}}{$\Hone$} &
                \multicolumn{1}{l@{\ \ $\to$}}{$\Hfour$} &
                \multicolumn{1}{l}{$\Hfouri$}
            \\
            $\NnegO$ & 0 & 1 & 2 \\
            $\Nneg$ & 0 & 1 & 2,4 \\
            \hline
            MOTT &
                \multicolumn{1}{l@{\ \ $\to$}}{$\Htwo$} &
                \multicolumn{1}{l@{\ \ $\to$}}{$\Hthree$} &
                \multicolumn{1}{l}{$\Hthreei$}
            \\
            $\NnegO$ & 0 & 1 & 2 \\
            $\Nneg$ & 0 & 1 & 2,4 \\
            \hline
            MOTT &
                \multicolumn{1}{l@{\ \ $\to$}}{$\Hout$} &
                \multicolumn{1}{l@{\ \ $\to$}}{$\Hin$} &
                \multicolumn{1}{l@{\ \ $\to$}}{$\Hini$} &
                \multicolumn{1}{l}{$\Hinii$}
            \\
            $\NnegO$ & 0 & 1 & 2 & 3 \\
            $\Nneg$ & 0 & 1,3 & 2,4 & 3,5,7,9 \\
            \hline
            MOTT & & &
                \multicolumn{1}{l@{\ \ $\to$}}{$\Hfive$} &
                \multicolumn{1}{l}{$\Hfivei$}
            \\
            $\NnegO$ &   & & 2 & 3 \\
            $\Nneg$ &    & & 2 & 3,5,7 \\
            \hline
            \hline
            MOTT &
                $\HoneA$ & $\HoneB$ & $\HoneC$ & $\HoneD$
            \\
            $\NnegO$ & 1 & 1   & 2   & 2 \\
            $\Nneg$  & 3 & 1,3 & 4,6 & 4,6 \\
            \hline
            MOTT &
                $\HtwoA$ & $\HtwoB$ & $\HtwoC$ & $\HtwoD$
            \\
            $\NnegO$ & 1   & 1 & 2 & 2 \\
            $\Nneg$  & 1,3 & 1 & 4 & 4 \\
        \end{tabular}
    \end{ruledtabular}
\end{table}

\begin{figure*}
    \includegraphics[width=0.5\linewidth]{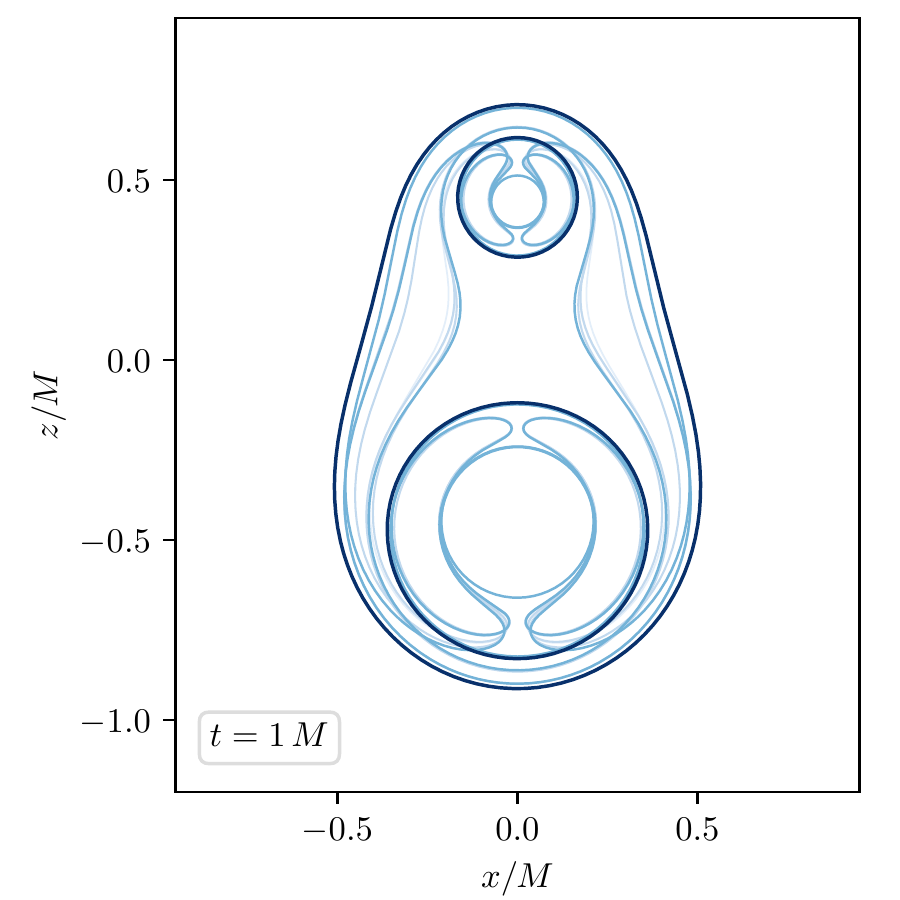}%
    \hfill%
    \includegraphics[width=0.5\linewidth]{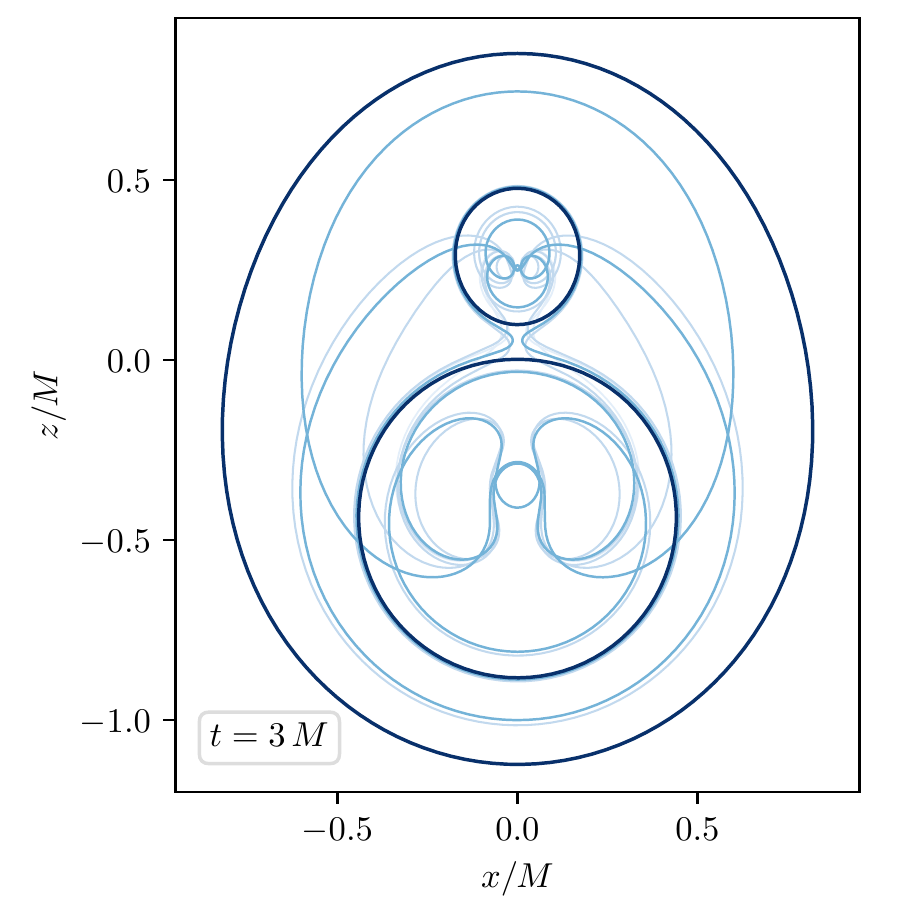}%
    \caption[]{\label{fig:MOTS_shapes_with_stability}%
        \MOTSs at two different times of the simulation. The line thickness
        and color reflects $\NnegO$, i.e. the number of negative stability
        eigenvalues of the $m=0$ mode.
        The three dark lines in both panels are $\Sout$ (common), $\Sone$
        (upper) and $\Stwo$ (lower).
        Lighter colors show \MOTSs with larger $\NnegO$.
        Note that none of the \MOTSs extends beyond $\Sout$.
    }
\end{figure*}

An interesting observation we can make here is that the number of negative
eigenvalues necessarily changes as we follow a smooth MOTT across such a
bifurcation or annihilation.
Three of the \MOTSs, namely $\Sout$, $\Sone$, and $\Stwo$, possess a
positive principle eigenvalue, i.e. they are
strictly stable and
thus act as barrier for trapped and untrapped surfaces in a neighbourhood.
By our terminology, they are apparent horizons and the world tubes they trace
out are dynamical apparent horizons.
The above properties are what one would usually expect from horizons
associated with black holes.
All other \MOTSs we found possess one or more negative eigenvalues.
At any given time, all \MOTSs with a negative eigenvalue are contained in the
interior of one of the strictly stable ones, i.e.
$\Sone$, $\Stwo$, or $\Sout$.
To present our results more systematically, let $\Nneg$ be the number of
eigenvalues $\lambda_{l,m} < 0$ and $\NnegO$ the number of eigenvalues
$\lambda_{l,m=0} < 0$.
Table~\ref{tab:Nneg} lists the various values of $\Nneg$ and $\NnegO$ we find
for the \MOTSs along each MOTT during the evolution
and FIG.~\ref{fig:MOTS_shapes_with_stability} shows all \MOTSs at two
different times with $\NnegO$ indicated by line thickness and color.
In each instance where a MOTS transitions through a bifurcation or
annihilation along the indicated sequences of \MOTTs, one additional negative
eigenvalue of the $m=0$ mode appears.

Another observation is related to the non-smooth MOTS mergers where
two \MOTSs touch at one point and coincide at this time with a MOTS
having a cusp. We were able to explicitly resolve three of
these mergers numerically, namely
$\Sone\cup\Stwo = \Sin$,
$\SoneC\cup\Stwo = \Sinii$ and
$\Sone\cup\StwoA = \Sfouri$.
Based on our results, we expect at least two more such mergers,
which we could not resolve for numerical reasons. These are
$\SoneA\cup\Stwo = \Sthreei$ and
$\SoneA\cup\StwoA = \Sfivei$.
For all these cases where $\Surf\cup\Surf'=\Surf''$, we find, with obvious
notation, that
\begin{equation}\label{eq:mergerlaw}
    \NnegO+{\NnegO}'+1 = {\NnegO}'' \,.
\end{equation}

\begin{figure}
    \includegraphics[width=0.9\linewidth]{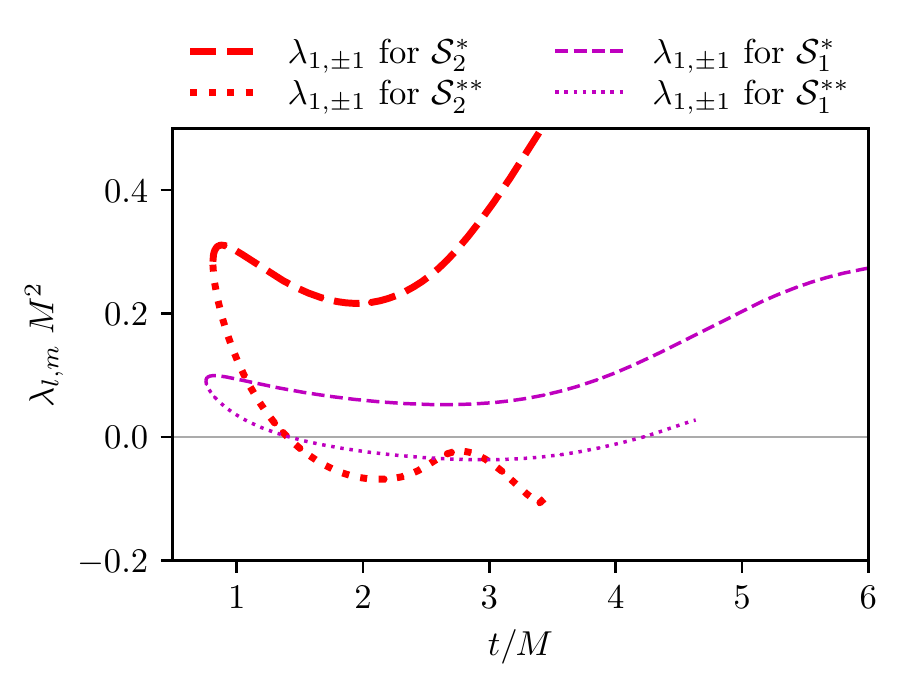}%
    \caption{\label{fig:BL31_S3i_stability_zero_crossings}%
        Examples of eigenvalues with $m\neq0$ crossing zero on the smoothly
        evolving portions of $\Sthreei$ (thick dotted line) and $\Sfouri$
        (thin dotted line).
        The smoothly connecting respective curves for $\Sthree$ and $\Sfour$
        are added here for reference.
        Note that both $\Sthreei$ and $\Sfouri$ have a principal eigenvalue
        $\lambda_{0} < 0$ and $\lambda_{1,0} < 0$, which are not shown.
    }
\end{figure}

Note that $\NnegO$ is constant along each individual MOTT, even when cusps and
self-intersections form, as they do for $\Sin$, $\Sinii$ and $\Sfouri$.
However, in several instances, we find that eigenvalues of the higher angular
modes ($m\neq0$) do cross zero on perfectly smooth portions of the MOTT.
Due to the axisymmetry and absence of spin in our simulation, we have a
$\pm m$ degeneracy in the spectrum, whence all zero crossings of eigenvalues
$\lambda_{l,m\neq0}$ happen in multiples of $2$.
Two examples of such cases are depicted in
FIG.~\ref{fig:BL31_S3i_stability_zero_crossings}, which shows that
the two degenerate eigenvalues $\lambda_{1,\pm1}$ of
$\Sthreei$ and $\Sfouri$ do cross zero during their evolution.
This crossing happens twice for the latter case.
Taking invertibility of $\Lslice$ as indicator for the existence of a smooth
evolution of a MOTS $\Surf$, we here have
explicit counterexamples showing that the converse of this statement is not
true. In other words, invertibility of $\Lslice$ is only a sufficient but not
a necessary condition for a smooth evolution.

\section{Signature and ingoing expansion}
\label{sec:signature}

\begin{figure*}
    \begin{tikzpicture}
        \node[anchor=south west,inner sep=0] (tube1) at (0, 0) {%
            \includegraphics[trim=-220 0 -220 -80,width=0.5\linewidth]{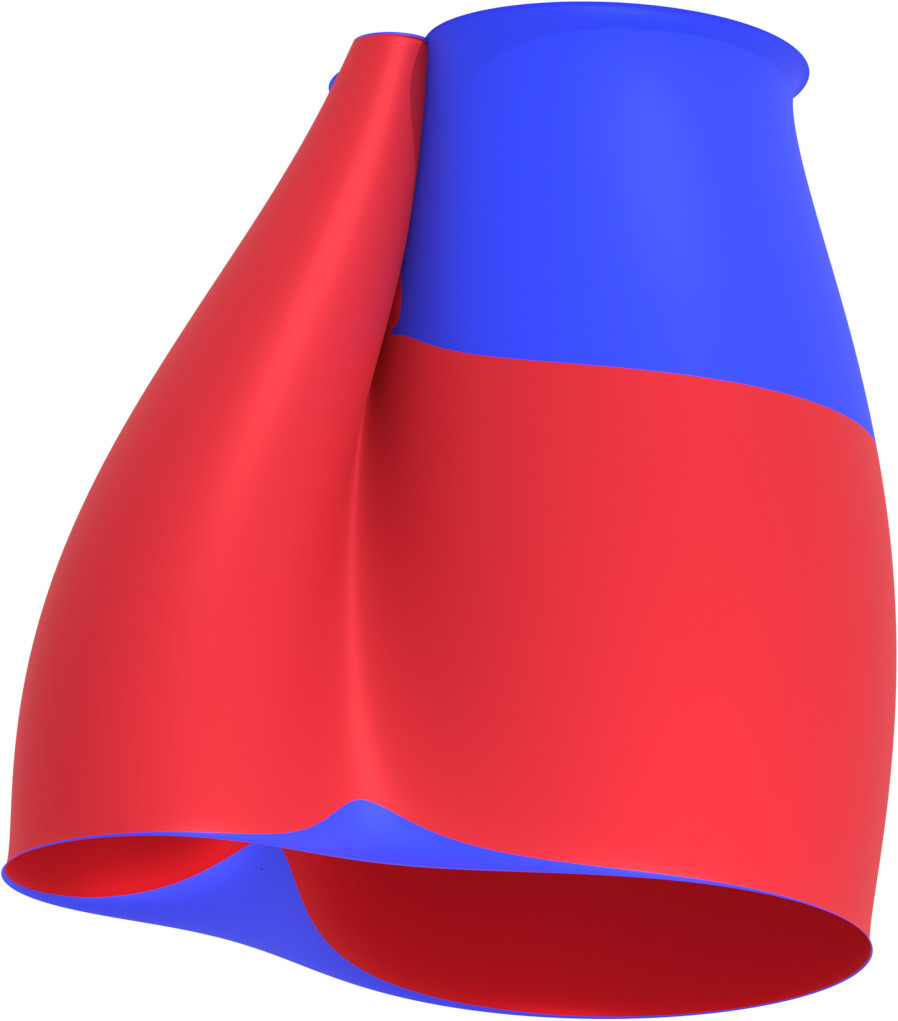}
        };
        \begin{scope}[x={(tube1.south east)},y={(tube1.north west)}]
            \node[anchor=north, opacity=0.5,text opacity=1,fill=white,inner sep=0.5ex] at (0.5,1.0) {
                $\Hin$
            };
        \end{scope}
        \node[anchor=west,inner sep=0] (tube2) at (tube1.east) {%
            \includegraphics[trim=-180 0 -180 -80,width=0.5\linewidth]{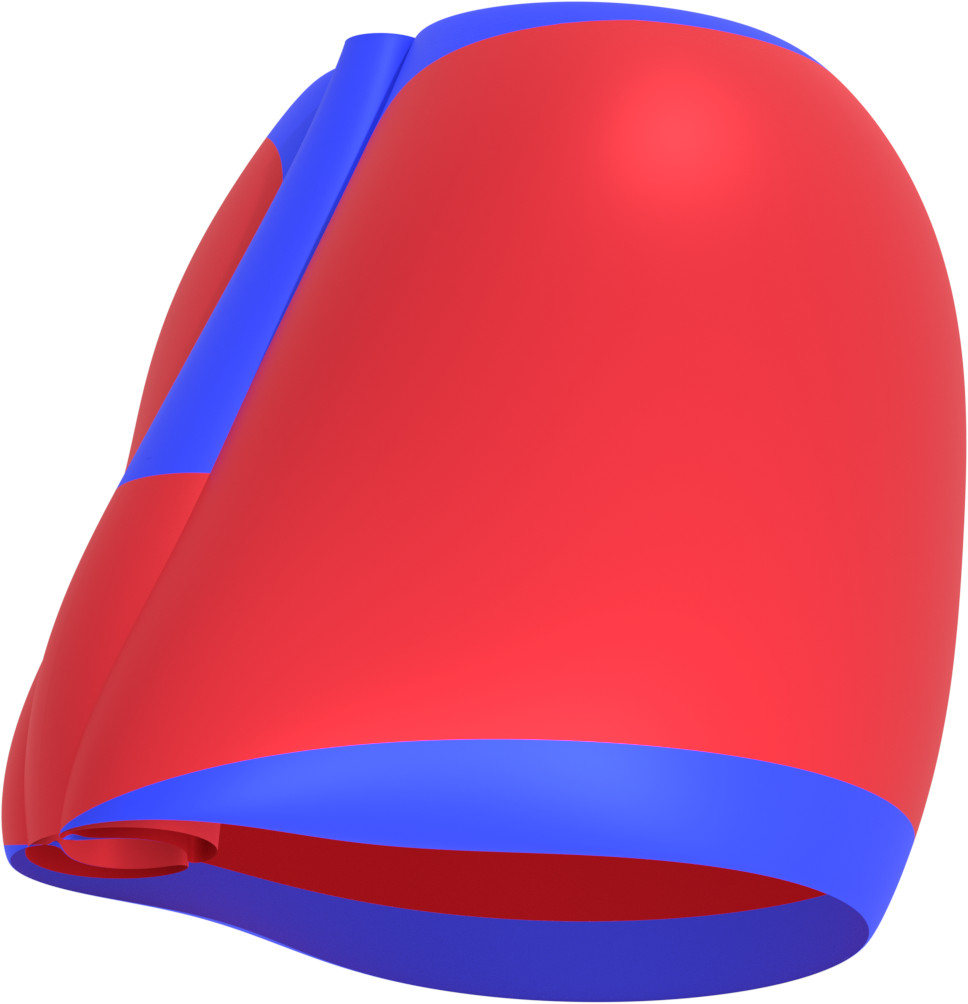}%
        };
        \begin{scope}[x={(tube2.south west)},y={(tube2.north west)}]
            \node[anchor=north, opacity=0.5,text opacity=1,fill=white,inner sep=0.5ex] at (0.5,1.0) {
                $\Hini$
            };
        \end{scope}
        \node[anchor=north] at ($(tube1.north)!0.5!(tube2.north)$) {%
            \includegraphics[scale=0.85]{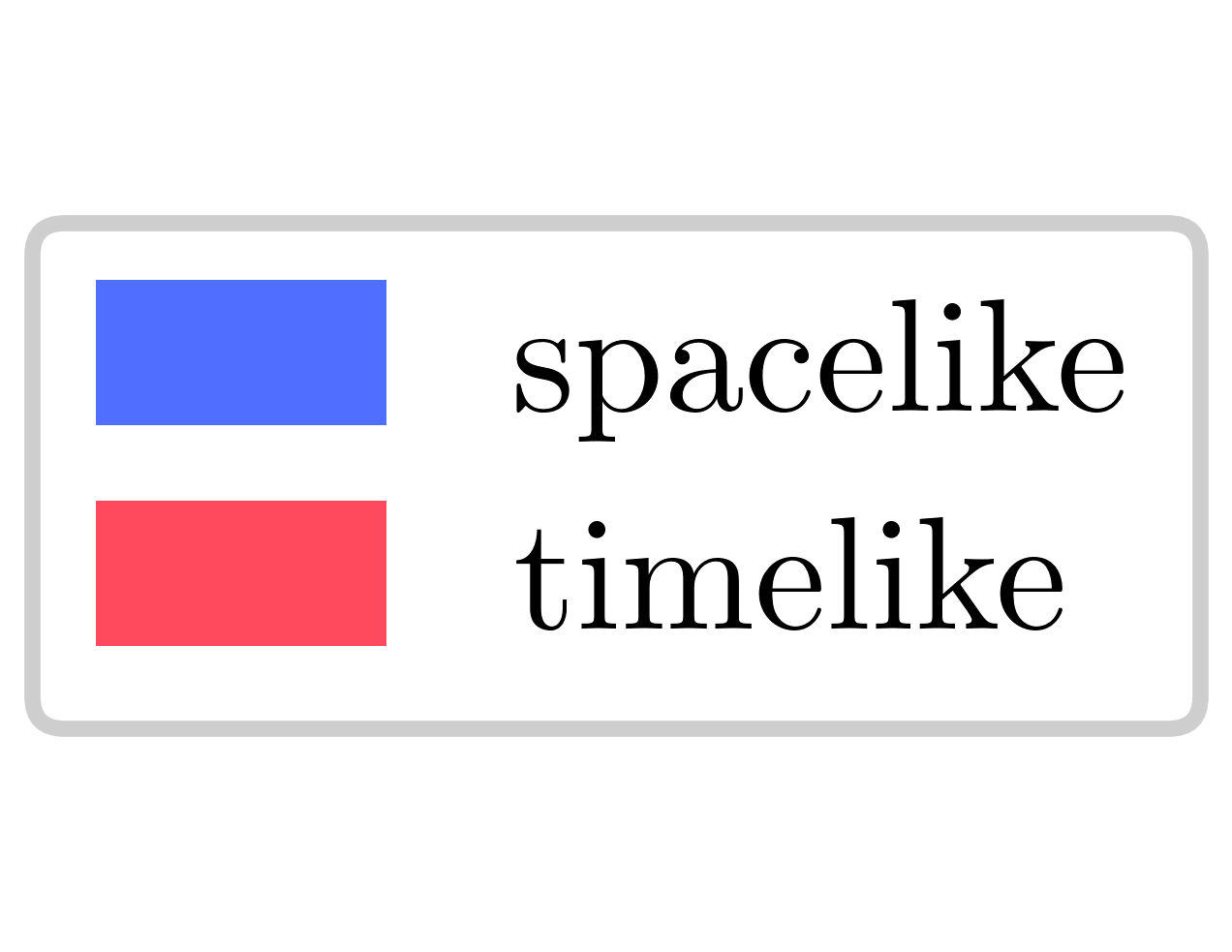}%
        };
    \end{tikzpicture}
    \caption{\label{fig:BL31_in_signature}%
        Signature of $\Hin$ (left panel) and $\Hini$ (right panel).
        In these tube plots, time goes upwards and $z$-values increase to the
        left. These two \MOTTs smoothly connect at their top ends but not at
        the bottom as they connect with different world tubes there.
        The bottom end of the right panel corresponds to the bottom-left panel
        of FIG.~\ref{fig:evolution_outer}.
    }
\end{figure*}

\begin{figure*}
    \begin{tikzpicture}
        \node[anchor=south west,inner sep=0] (tube1) at (0, 0) {%
            \includegraphics[trim=-220 0 -220 -80,width=0.5\linewidth]{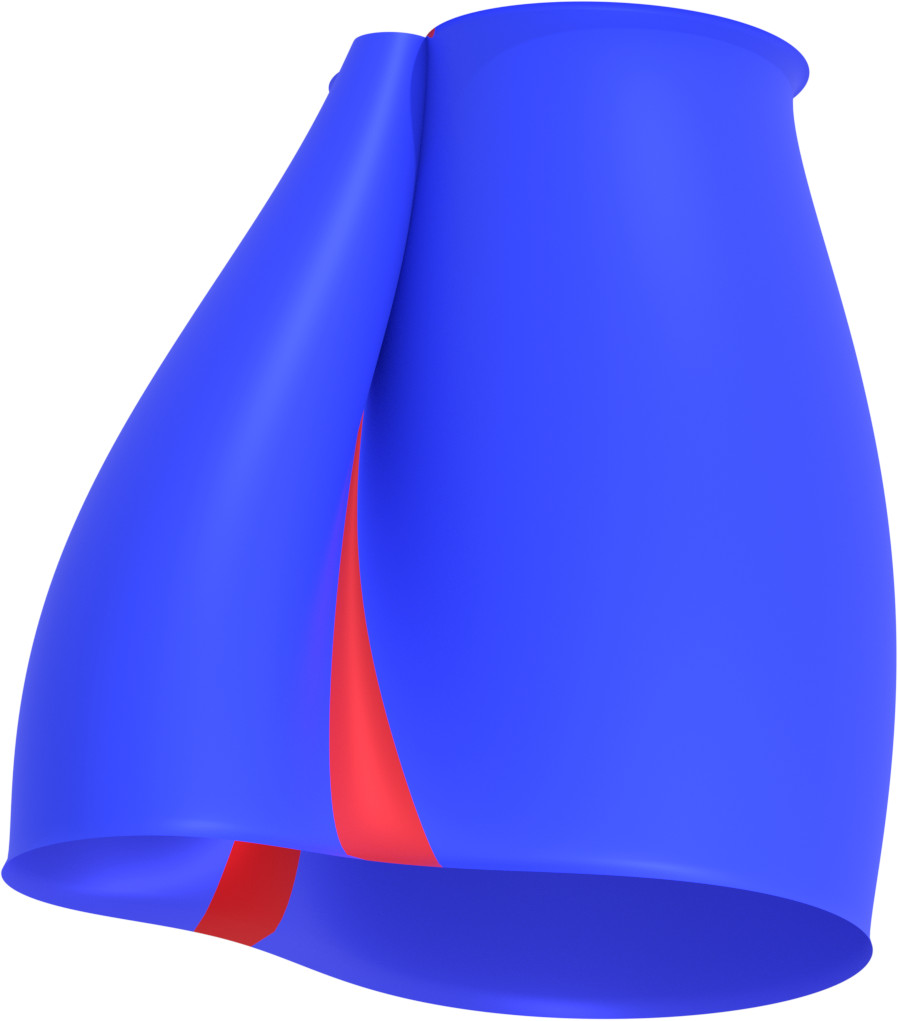}%
        };
        \begin{scope}[x={(tube1.south east)},y={(tube1.north west)}]
            \node[anchor=north, opacity=0.5,text opacity=1,fill=white,inner sep=0.5ex] at (0.5,1.0) {
                $\Hin$
            };
        \end{scope}
        \node[anchor=west,inner sep=0] (tube2) at (tube1.east) {%
            \includegraphics[trim=-180 0 -180 -80,width=0.5\linewidth]{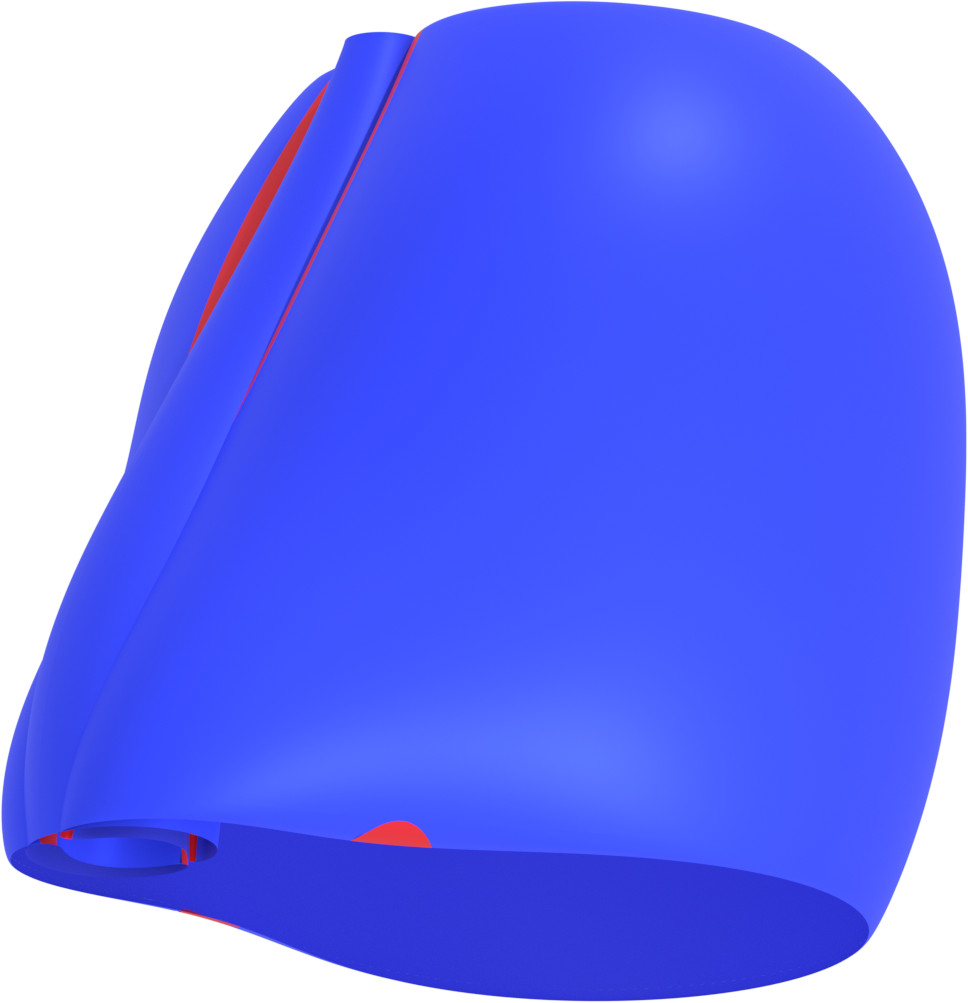}%
        };
        \begin{scope}[x={(tube2.south west)},y={(tube2.north west)}]
            \node[anchor=north, opacity=0.5,text opacity=1,fill=white,inner sep=0.5ex] at (0.5,1.0) {
                $\Hini$
            };
        \end{scope}
        \node[anchor=north] at ($(tube1.north)!0.5!(tube2.north)$) {%
            \includegraphics[scale=0.85]{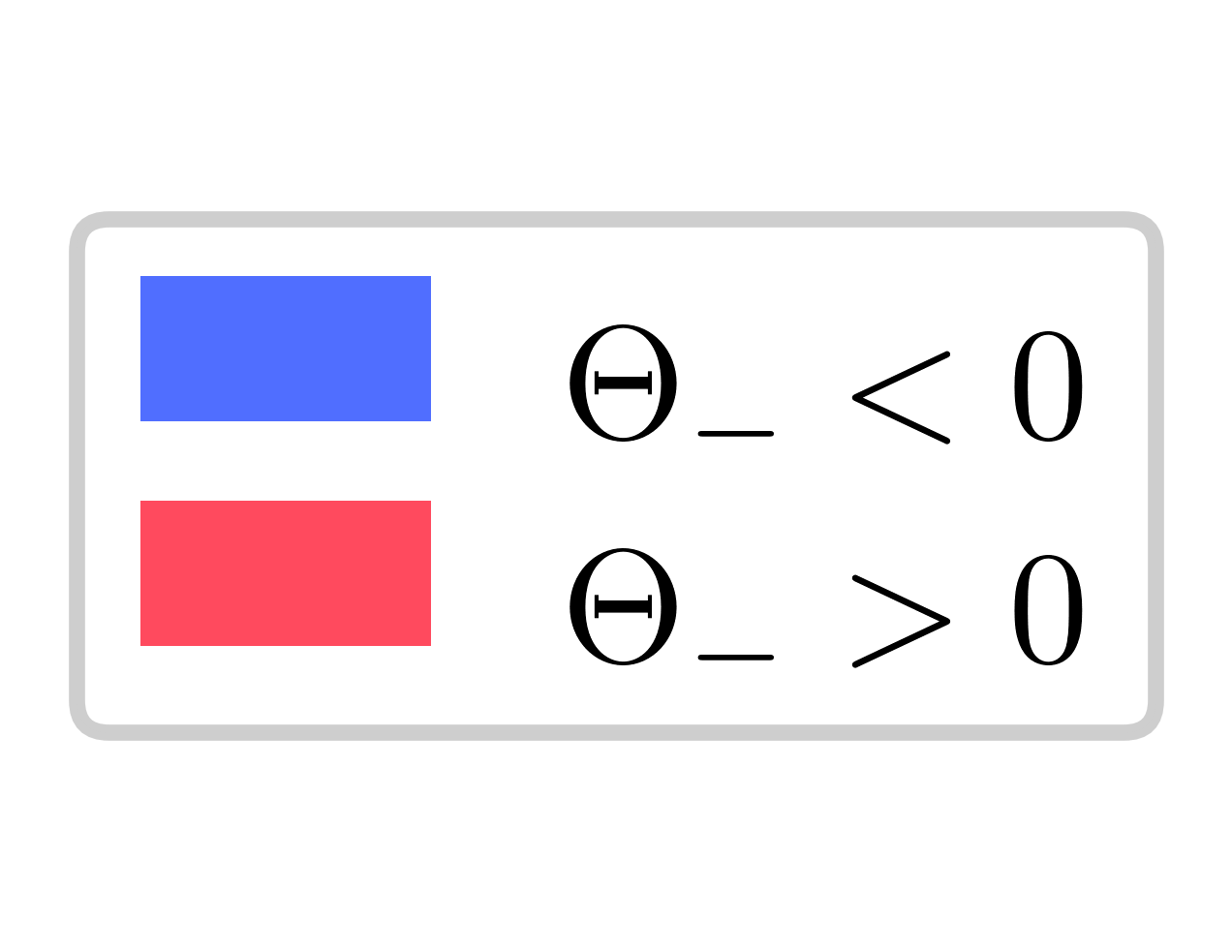}%
        };
    \end{tikzpicture}
    \caption{\label{fig:BL31_in_ingoing_exp}%
        Sign of the ingoing expansion $\Theta_{-}$ of $\Sin$ (left panel) and
        $\Sini$ (right panel) plotted on their respective \MOTTs $\Hin$ and
        $\Hini$ in the same perspective as in FIG.~\ref{fig:BL31_in_signature}.
        The portion with $\Theta_{-} > 0$ of $\Sin$ (left panel) smoothly
        connects to a corresponding portion on $\Sout$, which quickly vanishes
        after $\Delta\tname \approx 0.04M$ (not shown here).
    }
\end{figure*}

\begin{figure*}
    \begin{tikzpicture}
        \node[anchor=south west,inner sep=0] (tube1) at (0, 0) {%
            \includegraphics[trim=120 0 320 -80,clip,width=0.45\linewidth]{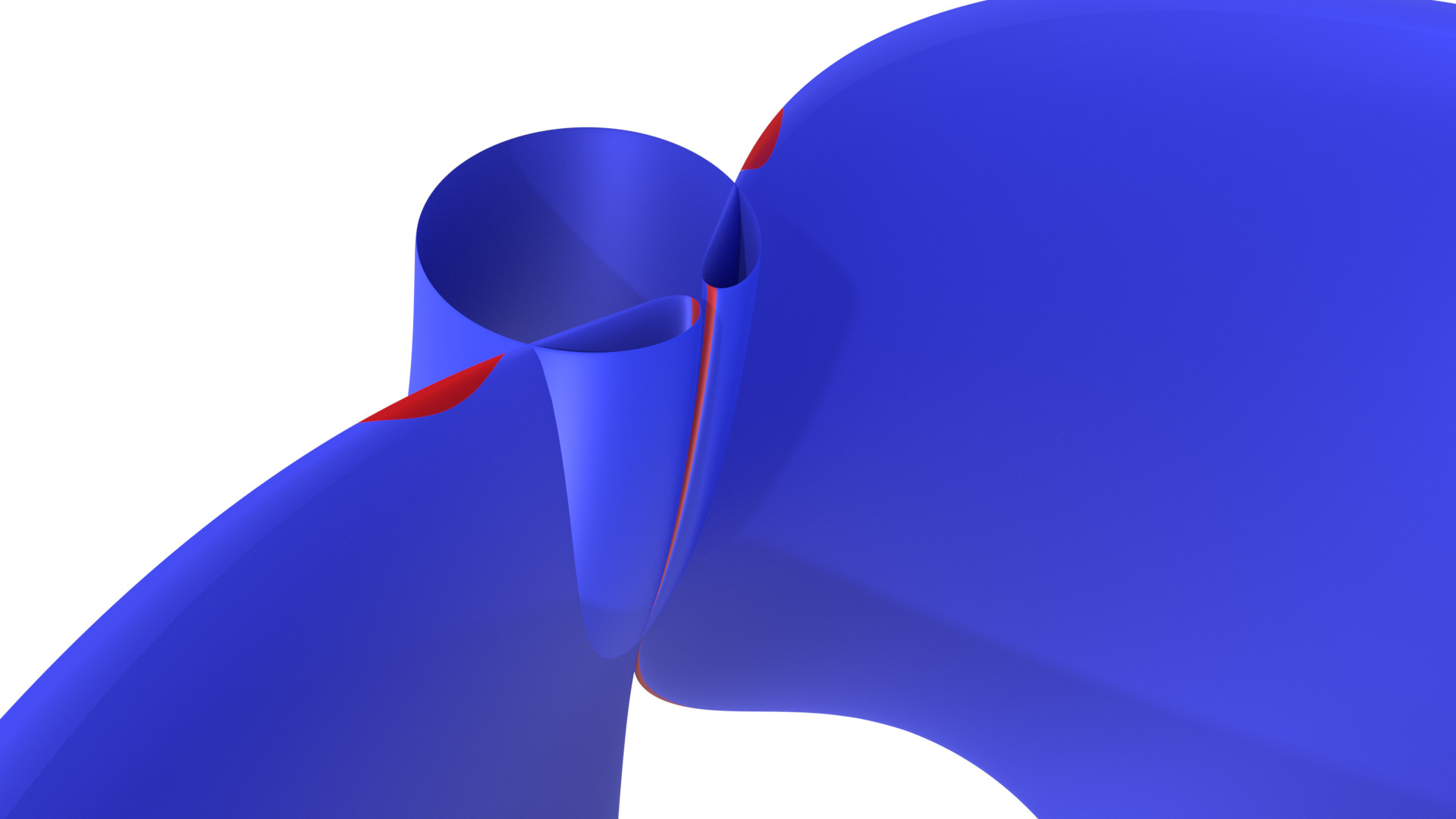}%
        };
        \begin{scope}[x={(tube1.south east)},y={(tube1.north west)}]
            \node[anchor=north, opacity=0.5,text opacity=1,fill=white,inner sep=0.5ex] at (0.5,1.0) {
                $\Hin$
            };
        \end{scope}
        \node[anchor=west,inner sep=0] (tube2) at ($(tube1.east)+(0.05\linewidth,0)$) {%
            \includegraphics[trim=120 0 320 -80,clip,width=0.45\linewidth]{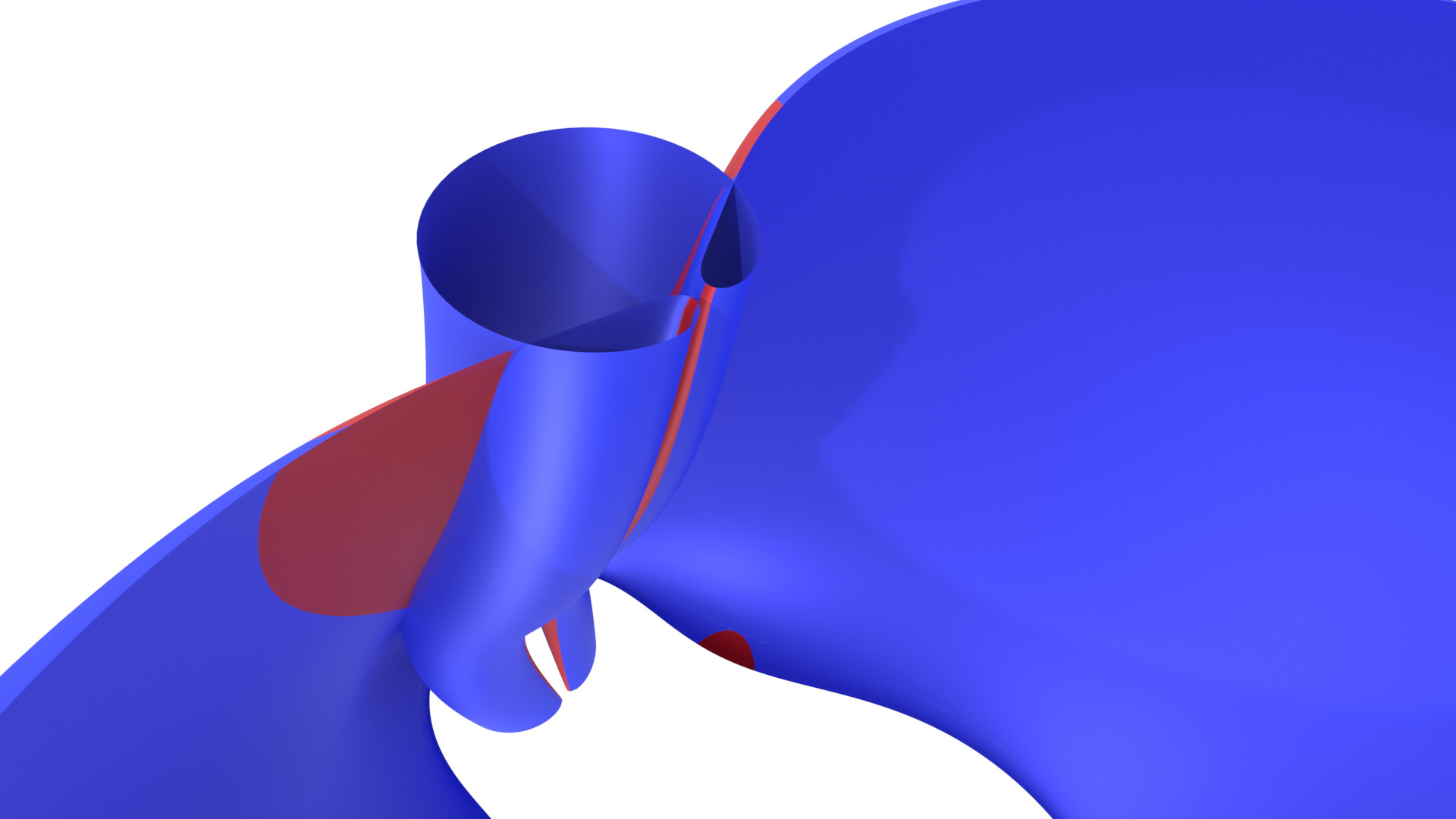}%
        };
        \begin{scope}[x={(tube2.south west)},y={(tube2.north west)}]
            \node[anchor=north, opacity=0.5,text opacity=1,fill=white,inner sep=0.5ex] at (0.5,1.0) {
                $\Hini$
            };
        \end{scope}
        \node[yshift=1em,anchor=north west] at (tube1.north west) {%
            \includegraphics[scale=0.85]{figs/exp_legend}%
        };
    \end{tikzpicture}
    \caption{\label{fig:BL31_in_ingoing_exp_top}%
        Close-up of the top ends of the \MOTTs shown in
        FIG.~\ref{fig:BL31_in_ingoing_exp}.
        The two world tubes smoothly join at this top end and the portions with
        $\Theta_{-} < 0$ connect across this merger.
    }
\end{figure*}

As discussed in Section~\ref{sub:stabilityGeneral}, a strictly
stable MOTS, $\lambda_0 > 0$, belongs to a dynamical apparent horizon that has spacelike
signature at that point (cf. \cite{Andersson:2005gq, Andersson:2007fh}).
Together with $\Theta_{-} \leq 0$, the area will be non-decreasing.
Since we see in FIG.~\ref{fig:BL31_areas_overview} that many of the \MOTSs
have a decreasing or non-monotonic area evolution, we expect that those
with $\lambda_0 < 0$ cannot 
have both non-positive ingoing
expansion $\Theta_{-} \leq 0$
and evolve along a spacelike MOTT.
Examples are shown in Figures~\ref{fig:BL31_in_signature} and
\ref{fig:BL31_in_ingoing_exp} where we see complicated signature changes and
indefiniteness of the sign of $\Theta_{-}$ along the world tubes $\Hin$ and
$\Hini$.
In these figures, time increases upwards and the signature or sign of
$\Theta_{-}$ is shown as color on the world tubes.
FIG.~\ref{fig:BL31_in_ingoing_exp_top} shows a close-up of the sign of
$\Theta_{-}$ at the top end where $\Sin$ and $\Sini$ annihilate (or
equivalently where $\Hin$ and $\Hini$ connect smoothly).
A qualitatively very similar behavior of these quantities is found for all
\MOTTs except $\Hout$, $\Hone$ and $\Htwo$.
These are purely spacelike and have $\Theta_{-} \leq 0$ at most
times.\footnote{%
    We do find a very short duration of $\Delta\tname \approx 0.04M$ after $\tAH$
    where $\Sout$ has a small portion with $\Theta_{-} > 0$ close to its
    equator. This was also found in \cite{pook-kolb2020II}.
    A similar portion with $\Theta_{-} > 0$ is found on $\Stwo$ shortly
    ($\Delta\tname \approx 0.02M$) before it annihilates with $\Sthree$.
    Both of these portions smoothly connect with corresponding portions on the
    \MOTSs they connect to ($\Sin$ for $\Sout$ and $\Sthree$ for $\Stwo$).
}

A result of Bousso and Engelhardt \cite{Bousso:2015mqa,Bousso:2015qqa} shows
that even when $\HH$ changes direction in time and has non-spacelike
segments, it will have a monotonic area evolution provided several conditions
hold on $\HH$.
One of these is that $\Theta_{-} \leq 0$, which we have already seen to not be
satisfied for most \MOTTs we found in this simulation.
At this point, one could immediately conclude that this result is not
applicable to most of our cases and hence finding non-monotonic area evolutions
is  not in tension with any theoretical expectation.
While certainly true, we still think it is worthwhile to show which of the
other assumptions made in the proof are violated, not least because they were
believed to be unrestrictive.

To state the relevant ones here, we go back to the evolution vector
$\mathcal{V}^\alpha$ defined as tangent to $\HH$ and orthogonal to each MOTS $\Surf$
foliating $\HH$.
As before, we fix its scaling by requiring
$\mathcal{L}_\mathcal{V} t = 1$.
Since the null normals $\ell^\pm$ span the two-dimensional space of normals to
$\Surf$, we can write
\begin{equation}\label{eq:tev}
    \mathcal{V}^\alpha = b \ell_+^\alpha + c \ell_-^\alpha.
\end{equation}
As $\mathcal{V} \cdot \mathcal{V} = -2bc$, the coefficients $b$ and $c$ are related to the
signature of $\HH$, i.e. $\HH$ is spacelike, timelike, or null when
$bc < 0$, $bc > 0$, or $bc = 0$, respectively.
The proof in \cite{Bousso:2015qqa} now requires in addition to genericity
assumptions satisfied in all our cases, the following:
\begin{enumerate}
    \item[(i)] Every inextendible portion of definite sign of $c$ is entirely
        timelike or contains at least one full MOTS.
    \item[(ii)] Every MOTS in $\HH$ splits a Cauchy slice $\Sigma$ that  it is
        contained in into two disjoint portions.
\end{enumerate}
Then, without restrictions on $\Theta_{-}$, it is proven that $c$ cannot change
sign on $\HH$.
On all but the three strictly stable \MOTTs, however, we find that
at least one of the above conditions is violated and that, in fact,
both $b$ and $c$ do change sign.

In cases of self-intersections, it is clearly condition (ii) that
does not hold.
But even for \MOTSs that do not self-intersect (or on portions of their
world tubes on which they do not self-intersect), we find that condition (i)
is violated.
For the case of $\Sin$, this was discussed in great detail in \cite{pook-kolb2020I}.
With the annihilation of $\Sin$ with $\Sini$, we are now able to extend these
results to later times.
In particular, shortly before $\Sin$ vanishes, its world tube $\Hin$ becomes purely spacelike
with $c < 0$ on full \MOTSs.
At these times, however, $\Sin$ has self-intersections, i.e. this presents an
explicit example that (i) is not a sufficient condition.

\section{Conclusions}
\label{sec:conclusions}

In the present second paper of this two-part study, the new
generalized shooting method introduced in the first paper was used
successfully to uncover new \MOTSs forming during the head-on
merger of two non-spinning black holes, including \MOTSs of toroidal topology.
This has vastly increased the number and variety of known \MOTSs and also shows that they can have a 
much richer range of geometrical properties than had been previously expected. 

However this increase has also highlighted the rarity and significance of stable \MOTSs. Only three out of all the multitude
that we have observed are stable -- even strictly stable except for the points of annihilation or
bifurcation --
and trace out spacelike world tubes. These are exactly the \MOTSs that one would naturally associate
with black hole boundaries: the
two individual black holes ($\Sone$ and $\Stwo$) and the final
remnant ($\Sout$).
These world tubes $\Hone$, $\Htwo$ and $\Hout$ are the dynamical apparent horizons.
One may ask whether additional strictly stable \MOTSs may exist if
we do not restrict ourselves to only axisymmetric surfaces.
Fortunately, this has been ruled out by Theorem~8.1 in
Ref.~\cite{Andersson:2007fh}.
This unambiguous natural choice of physically relevant horizons provides an
additional numerical indication that dynamical apparent horizons are well-behaved
objects suitable to describe the highly dynamical and non-perturbative regime
during such a merger.

That said, the apparent horizons cannot forever remain aloof from the common herd. 
$\Sout$ appears out of a bifurcation with the unstable $\Sin$ while $\Stwo$ and (likely) $\Sone$ 
are ultimately annihilated in mergers with other unstable \MOTSs.
The additional \MOTTs then significantly increase our
understanding of the interior structure forming shortly after the common
apparent horizon appears.
This structure shares certain features with previous speculations that the
merger might, in fact, be described by a single smooth MOTT weaving back and
forth in time \cite{Hayward:2000ca,Booth:2005ng,Moesta:2015sga,Gupta:2018znn}.
What we find is significantly more complicated, but we do see this back and
forth in time.
Since we lose the \MOTSs only for (well understood) numerical reasons, it
seems plausible that they continue to weave back and forth, possibly
forming more and more self-intersections.
We also find that all world tubes seem to be connected\footnote{%
    The likely annihilation of $\Sone$ with $\Sfour$ is discussed in
    Section~\ref{sub:worldtube_S1}.
    Similarly, based on our results we expect mergers of
    $\Sthreei$ with $\Stwo \cup \SoneA$
    and
    $\Sfivei$ with $\SoneA \cup \StwoA$
    to happen via cusp-formation.
}
However, since no MOTS crosses a puncture, some connections are not smooth
and instead happen via cusp-formation with subsequent self-intersections.

Together, these observations motivate the following suggestion.
If one assumes that
(i) \MOTSs cannot cross punctures,
(ii) \MOTSs appear and disappear only as pairs,
(iii) all \MOTTs connect in some form and
(iv) the individual apparent horizons vanish at some point during a merger,
then many of the observed behaviors seem to be inevitable.
It seems conceivable that (i) holds and the results of Andersson et al.
\cite{Andersson:2008up} points toward (ii).
While our results certainly do not {\em imply} (iii) and (iv), they might
still provide an incentive for further investigation in this direction.

The present results for the axisymmetric head-on collision of two black holes
also have implications for the generic case where inspiraling black holes
coalesce without any symmetry.
We now know which kinds of surfaces a generalized MOTS finder must be able to
resolve and which surfaces to look for.
One might explore whether a generalization of the shooting method
could be used to approximate near-axisymmetric \MOTSs to be used as initial
guesses for such a finder.
An important question will be whether it is still only three \MOTTs which
are stable and hence dynamical apparent horizons.

\begin{acknowledgments}
     We would like to express our gratitude to 
    Graham Cox, 
    Jose~Luis~Jaramillo,
    Badri~Krishnan, Hari Kunduri and the members of the 
    Memorial University Gravity Journal Club
    for valuable discussions and suggestions.
 
   IB was supported by the Natural Science and Engineering Research Council of Canada Discovery Grant 2018-0473. 
   The work of RAH was supported by the Natural Science and Engineering Research Council 
   of Canada through the Banting Postdoctoral Fellowship program and also by AOARD Grant FA2386-19-1-4077.
\end{acknowledgments}

\appendix

\section{Families of surfaces of constant expansion}
\label{app:CESs}

A MOTS $\Surf$ is defined as a closed surface with zero outward expansion,
$\Theta_{+} = 0$.
However, one may also try to look for surfaces $\Surf_c$ with
$\Theta_{+} = c$, where $c = \text{const}$, in a neighbourhood of any given MOTS.
These surfaces can be used to help locating common \MOTSs as early as possible
during the simulation by tracking common surfaces $\Surf_{c>0}$, which are
seen to exist prior to the formation of a common MOTS $\Surf = \Surf_{c=0}$
\cite{Schnetter:2003pv,Schnetter:2004mc}.
We will show here another application of such surfaces, which turned out to be
helpful in resolving the various bifurcations and annihilations.
This is related to and motivated by the observation made in
\cite{pook-kolb:2018igu} that a family of such surfaces may connect one MOTS
with another. See Figure~13 in \cite{pook-kolb:2018igu} and its discussion for
details.

\begin{figure}
    \includegraphics[width=0.9\linewidth]{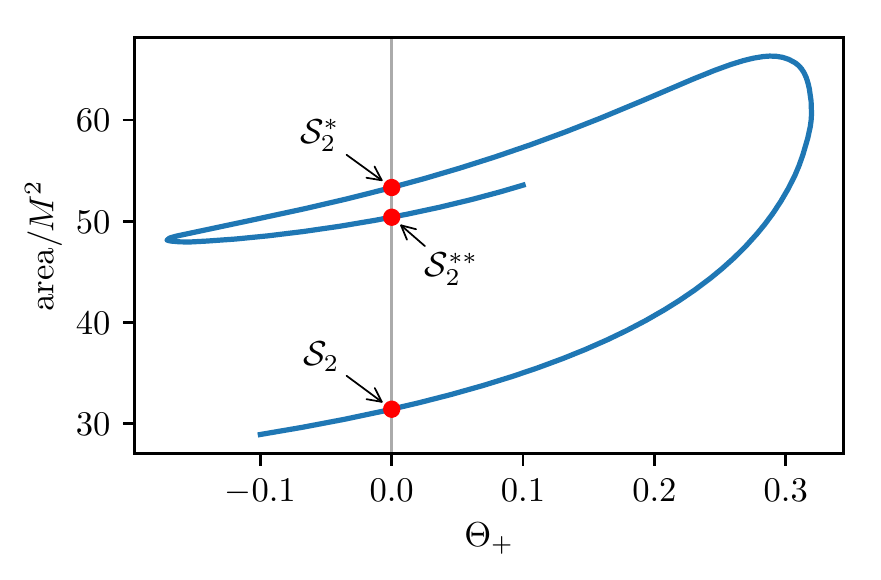}%
    \caption{\label{fig:BL31_CESs_area}%
        A family of surfaces of constant expansion shown in the plane of area
        and expansion.
        Each point on the solid line corresponds to one surface of this
        family.
        Whenever the curve crosses $\Theta_{+} = 0$, the surface is a MOTS.
        This plot shows a family constructed from $\Stwo$ at simulation time
        $\tname = 2.5M$, which connects $\Stwo$ with $\Sthree$ and $\Sthreei$.
    }
\end{figure}

\begin{figure*}
    \includegraphics[width=0.5\linewidth]{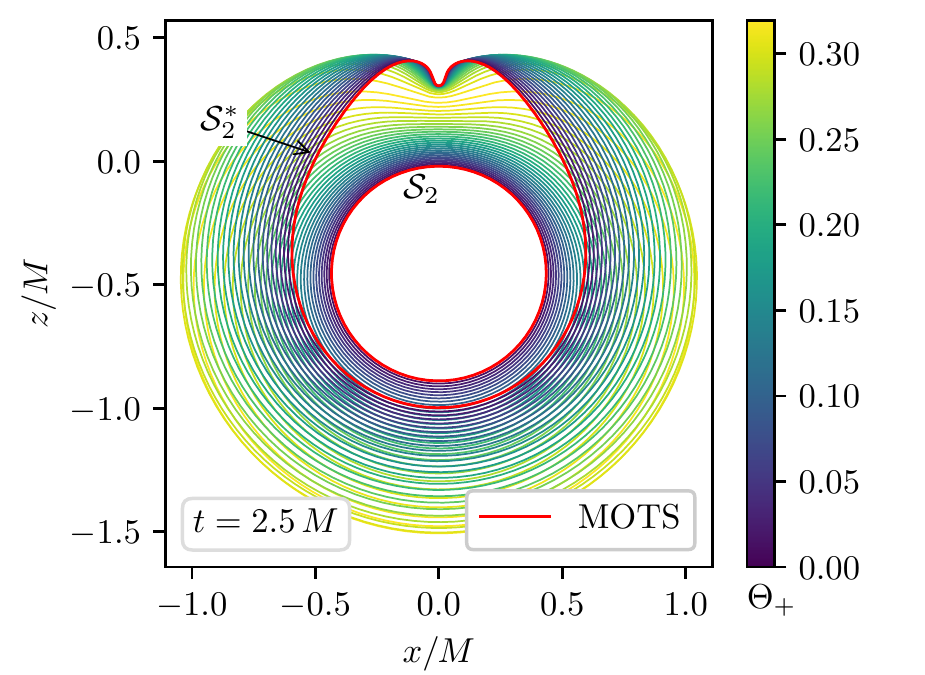}%
    \hfill%
    \includegraphics[width=0.5\linewidth]{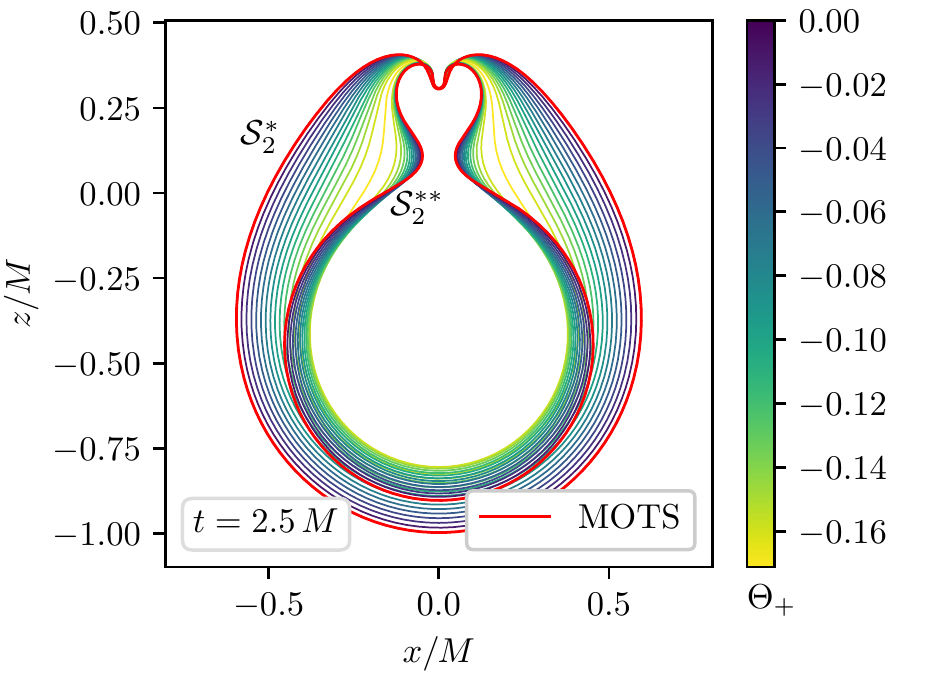}%
    \caption{\label{fig:BL31_CESs_S2_S3}%
        Shapes of the family of constant expansion surfaces of
        FIG.~\ref{fig:BL31_CESs_area}. Shown is the subset of surfaces
        connecting $\Stwo$ with $\Sthree$ (left panel)
        and the subset
        connecting $\Sthree$ with $\Sthreei$ (right panel).
        The color indicates the value of $\Theta_{+} = c$ while the \MOTSs
        are shown as solid red lines.
    }
\end{figure*}

We start with a MOTS $\Surf$ found in some particular Cauchy slice $\Sigma_t$
at simulation time $\tname = t_1$.
This surface is then tracked through the simulation forwards and backwards in
time.
If in either direction, the MOTS is lost and cannot be located anymore, say at
$t_2 > t_1$, then we choose a time $\tname \lesssim \tname_2$ and construct a
family of surfaces $\Surf_c$ starting with $\Surf_0 = \Surf$.
Note that just as there may be multiple \MOTSs $\Surf_{0}$ in any given Cauchy
slice, the surfaces $\Surf_c$ for $c\neq0$ will also not be unique in general.
However, when varying $c$ in small steps $c \to c' = c + \varepsilon$,
one can look for $\Surf_{c'}$ in the vicinity of $\Surf_c$ by taking $\Surf_c$
as initial guess.
As an example, FIG.~\ref{fig:BL31_CESs_area} shows such a family in terms of
the expansion and area of the $\Surf_c$.
In this case, we start from $\Stwo$ at a time $\tname = 2.5M$ and we are able
to reliably locate $\Sthree$ and $\Sthreei$.
FIG.~\ref{fig:BL31_CESs_S2_S3}
shows the shapes of these surfaces of constant expansion and how they connect
$\Stwo$ with $\Sthree$ (left panel) and
$\Sthree$ with $\Sthreei$ (right panel).

A slight complication is encountered whenever $\Theta_{+} = c$ has a local
extremum. This happens twice in FIG.~\ref{fig:BL31_CESs_area}.
In these cases we cannot vary $c$ but instead we can take a small step in area
by prescribing the area instead of the expansion for this step
(see e.g. \cite{Schnetter:2003pv}).
Alternatively, one can {\em anticipate} the shape change by extrapolating from
the previous steps to construct an initial guess surface which overcomes the
extremum of $c$.

Once another MOTS is found this way, it can itself be tracked forwards and
backwards in time to resolve the possible annihilation or bifurcation.

\section{Annihilation of $\Sone$ with $\Sfour$}
\label{app:S1annihilation}

\begin{figure*}
    \includegraphics[width=0.45\linewidth]{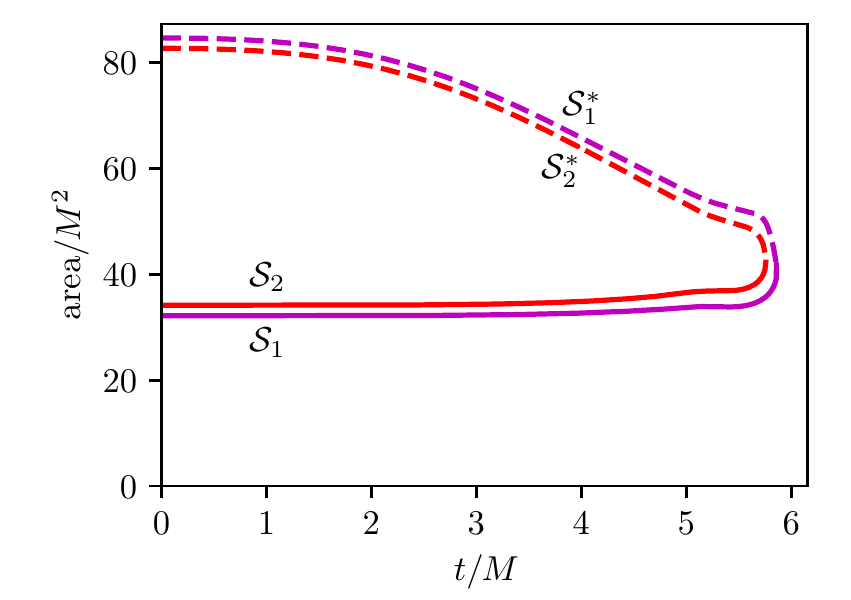}%
    \hfill%
    \includegraphics[width=0.45\linewidth]{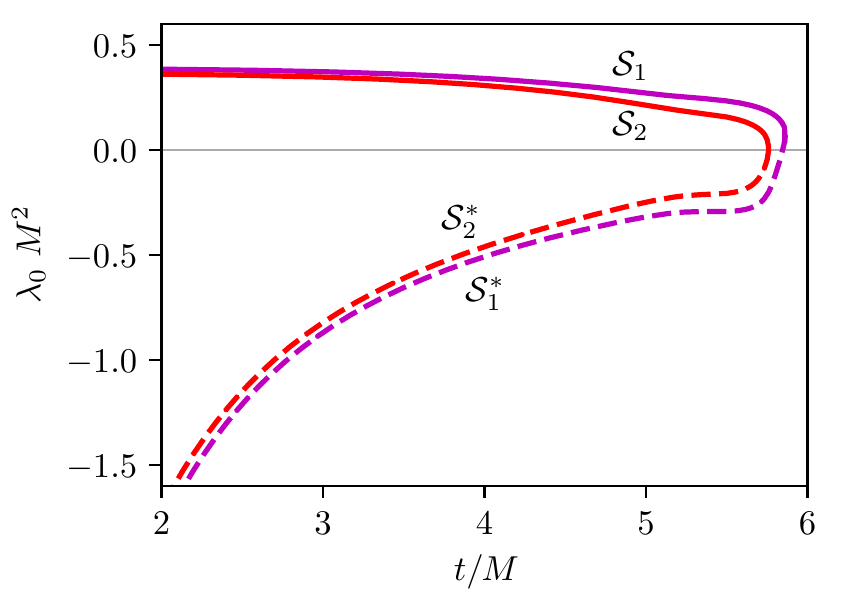}%
    \caption[]{\label{fig:S1_annihilation}%
        Evolution of the area (left panel) and principal stability
        eigenvalue (right panel) for a simulation with initial
        conditions $q=m_2/m_1=1.05$ and $d=0.4$.
        We focus here on the individual apparent horizons of $\Sonetwo$ and the
        \MOTSs with which they annihilate, $\Sfour$ and $\Sthree$,
        respectively.
        These plots show a simulation with resolution
        $1/\Delta x = 960$. Note that the final part for
        $\tname\gtrsim5M$ still suffers from the \MOTSs being too
        close to the numerically underresolved puncture regions.
    }
\end{figure*}

The annihilation of the larger individual MOTS $\Stwo$ with $\Sthree$ was
found with high accuracy and is discussed in the main text.
Here, the goal is to show that it is plausible that the smaller individual
MOTS $\Sone$ also annihilates, in this case with $\Sfour$, and that it does
not have a qualitatively different behavior than $\Stwo$ in this regard.
To this end, we perform a simulation with different initial conditions than
those in the main text and show that both individual horizons annihilate.
We take this as suggesting that the lack of annihilation in the main
configuration is purely due to the numerical setup and caused by the MOTS
moving too close to the puncture (in the numerical coordinates) as also
analyzed in \cite{evans:2020lbq}.

By using more equal masses, the shape of the smaller apparent horizon
remains larger in coordinates for a longer time, and a smaller initial
distance parameter reduces the simulation time when the annihilation takes
place.
FIG.~\ref{fig:S1_annihilation} shows the area and principal
stability eigenvalue for a simulation with a mass ratio of $q=1.05$
and distance parameter $d=0.4$.
We see that both individual apparent horizons seem to annihilate, first the
larger one $\Stwo$ with $\Sthree$ and shortly after the smaller one
$\Sone$ with $\Sfour$.
Despite the above choice of parameters for this simulation, the
\MOTSs still approach the punctures (in coordinates) before they
vanish. We found the annihilations of both apparent horizons in simulations
with spatial resolutions of $1/\Delta x = 480$, $720$, and $960$.
However, the precise behavior of these curves in the final time span
after $\tname\sim5M$ varies between the resolutions.
While not as convincing as the remaining results we present, a
merger of $\Sone$ and $\Sfour$ seems at least plausible.

\bibliography{blmotos}{}

\end{document}